\documentclass[aps,prd,preprint,groupedaddress]{revtex4-1}

\usepackage{amsmath,amssymb,amsthm,mathrsfs,amsfonts,dsfont,hyperref}
\usepackage{multirow}

\DeclareMathOperator{\tr}{tr}

\begin{document}


\title{Masses of the lowest spin-0 and spin-1 meson nonets: explicit symmetry breaking effects}


\author{J. Morais}
\email[]{jorge.m.r.morais@gmail.com}
\thanks{}
\altaffiliation{}
\affiliation{CFisUC, Department of Physics, University of Coimbra, 3004-516 Coimbra, Portugal}

\author{B. Hiller}
\email[]{brigitte@teor.fis.uc.pt}
\altaffiliation{}
\affiliation{CFisUC, Department of Physics, University of Coimbra, 3004-516 Coimbra, Portugal}

\author{A. A. Osipov}
\email[]{osipov@nu.jinr.ru}
\altaffiliation{}
\affiliation{Joint Institute for Nuclear Research, Bogoliubov Laboratory of Theoretical Physics, 141980 Dubna, Russia}


\date{\today}

\begin{abstract}
We extend a known multi-quark three-flavor Lagrangian of the Nambu–-Jona-Lasinio type, which includes a set of effective interactions proportional to the current quark masses, to include the multi-quark interactions of vector and axial-vector types. It is shown that the mass spectrum of the four low-lying meson nonets are in agreement with current phenomenological expectations. The role of the new interactions is analyzed in detail. 
\end{abstract}

\pacs{11.30.Rd, 11.30.Qc, 12.39.Fe, 14.40.-n}
\keywords{QCD, Chiral transformations, Spontaneous symmetry breaking, Hadronic effective field theory, Axial-vector mesons}

\maketitle

\section{Introduction}

Since the formulation of the Nambu--Jona-Lasinio (NJL) model, the study of the non-perturbative QCD vacuum and low-energy phenomena on the basis of chiral symmetry and its dynamical breaking modelled through effective multi-fermion interactions has seen considerable development. The NJL model was originally formulated in terms of a single effective four-fermion vertex of nucleon fields \cite{NJL1,NJL2}. The rise of the quark model and, later, of QCD, has led to its reinterpretation in terms of colored quark fields and to its extension to three flavors, which in turn led to the inclusion of the six-quark $U_A\left(1\right)$ breaking 't Hooft determinant term \cite{tHooft76,Kunihiro88,Bernard88,Reinhardt88}. Chiral eight-quark interactions have been included in later studies \cite{Alkofer89,Burvenich2003,Osipov2006}, completing the set of effective vertices in the chiral Lagrangian which are relevant to the dynamical breaking of chiral symmetry in four dimensions \cite{Andrianov93}. It has been argued in \cite{Osipov2006,Osipov2007}, using arguments pertaining to $N_c$ counting (including, in particular, effective potential stability conditions), that the six- and eight-quark vertices constitute next-to-leading order terms in a hierarchy of multi-quark interactions, as opposed to the four-quark term which is the leading order contribution.

Chiral symmetry is only an approximate symmetry of strong interactions, being explicitly broken in QCD due to finite current quark masses. The explicit chiral-symmetry breaking is usually included in the NJL-type models through a canonical mass term of the Dirac fermion field. It is known, however, that the mass of the $s$ quark is too large for $SU(3)\times SU(3)$ symmetry to be very reliable at lowest order in chiral-symmetry breaking. Thus, one should consider the next-to-leading order (in quark masses) contributions. Fortunately, the $1/N_c$ hierarchy of chiral symmetric multi-quark effective vertices implies that the hierarchy of explicit symmetry-breaking terms may also exist \cite{model1,model2} (the QCD origin of the set of explicit symmetry breaking multi-quark interactions has been traced back recently in \cite{Braghin16}). The higher-order effects contribute to the effective quark masses and lead to the chiral symmetry-breaking meson interactions. This extension has been shown to improve the accuracy of estimates within the NJL type effective approach in describing the scalar and pseudoscalar meson spectra, as well as in yielding reasonable results for some strong and radiative meson decays \cite{model1,model2}. This extended version of the model has been also employed in the thermodynamic study of the chiral transition and of quark matter in \cite{SQM}. It has been further developed in order to account for isospin breaking effects in \cite{Osipov2016}.

Any description of strong interactions would be incomplete without an extension of the above mentioned ideas to spin-1 states, most notably vector and axial-vector mesons. This generalization, as it will be shown in the text, covers two aspects of the effective multi-quark interactions. First, we include all possible chiral-symmetric multi-quark interactions of spin-0 and spin-1 types up to and including the eight-quark local couplings. As a result, the Lagrangian contains not only conventional scalar- and vector-type four-quark interactions but includes also their mixture: the eight-quark interactions made from spin-0 and spin-1 chiral symmetric combinations. Second, we classify and include all explicit symmetry-breaking multi-quark interactions in the approximation considered. Such a description may be regarded as an effective Lagrangian approach at the level of multi-quark vertices. Despite the generality of this approach, the model has a large number of coupling constants which must be fixed from phenomenology. While this fact may be seen as a drawback of the approach, it should be noted that these parameters are less than arbitrary, obeying strict symmetry constraints which bind them together in  very specific ways such that their sheer number is not an \textit{a priori} guarantee that the relevant observables may be accurately fitted. At zero temperature and density, the model has a limited predictive power due to a degeneracy between certain parameter sets. Nevertheless, this degeneracy is lifted as we introduce finite temperature and density, leading to its exceptional relevance in the study of thermodynamic properties of strongly interacting quark matter at finite density, temperature, or in a strong magnetic background, supplying us with a more detailed picture of quark dynamics.
 
Let us give just a few motivations. It has been shown recently \cite{Benic2014} that the presence of vector modes has an important impact on the equation of state, making it stiffer. Indeed, in \cite{SQM}, the model's equation of state with only spin-0 modes has been shown to be too soft for describing recently observed compact stellar objects, whereas in \cite{Lee2013,Renan2016} the presence of effective interactions involving vector modes has been shown to be instrumental in this description. In these works, the strengths of the vector related interactions are kept as free parameters. This provides a motivation for the consistent inclusion of spin-1 states. Another interesting prospect of the inclusion of spin-1 states is motivated by the phenomenologically successful ideas of vector-meson dominance (VMD) and universal coupling of mesons to conserved currents \cite{Sakurai1960} (it is well-known that the NJL model supports both ideas \cite{Volkov84}). We have a unique opportunity to study the role of the full set of effective multi-quark terms in the description of strong and radiative decays of vector mesons both in the unsymmetric Nambu-Goldstone phase and in the symmeric Wigner-Weyl phase, including the important details of the explicit symmetry-breaking phenomenon. This comparison may supply us with useful information about the possible signals of chiral symmetry restoration in hot and dense matter (where the role of eight-quark interactions is more profound \cite{Osipov2007b,Kashiwa2007,Kashiwa2008, Osipov2008,Hiller2010,SQM}), giving us insights into the structure of the QCD phase diagram. There are also indications that the location of the critical end point is affected by the special role of the vector-channel interaction in the medium \cite{Fukushima08,Benic2014}.

There are several different approaches aimed at including spin-1 mesons in the effective chiral Lagrangian \cite{Weinberg68,Coleman69I,Coleman69II,Geffen69,Ueda69,Meissner88,Bando88,Gasser89}. These contain the ``nonlinear realization'' \cite{Weinberg68,Coleman69I,Coleman69II}, ``massive Yang-Mills'' \cite{Geffen69,Ueda69,Meissner88}, ``hidden-gauge'' \cite{Meissner88,Bando88}, and ``antisymmetric tensor-field'' \cite{Gasser89} formalisms. Despite the rather different forms of their Lagrangians, all of these approaches are in principal equivalent \cite{Ecker89,Birse96,Kampf07}. Each corresponds to a different choice of spin-1 fields and their transformations. This is illustrated rather well in NJL-type models, where there is considerable freedom in the choice of auxiliary fields in the vector and axial-vector channels \cite{Ebert86,Osipov96,Hiller00}. In the present work the physical spin-1 fields belong to the linear representation of the chiral group. This scheme leads to the most economical structure of the effective Lagrangian. A further simplification is related with the way we remove the $\pi a_1$ mixing term. We do this by a linearized shift in the definition of the axial-vector field. This transformation does not lead to chiral symmetry violations \cite{Osipov17}, although it changes the chiral transformation properties of the axial-vector and vector fields in the broken vacuum. Such diagonalization generates a minimum number of vertices in the effective Lagrangian. 

Another kind of approach which treats the low-lying axial vector nonet as dynamically generated meson-meson resonances is also employed in chiral effective models. Examples may be found in \cite{Roca2005} and \cite{GarciaRecio2011,Lutz2004,Terschlusen2012}, with the latter works interpreting scalar mesons as being also dynamically generated resonances. These approaches are in contrast with other works such as \cite{Andrianov99,Peris2001}, where the four spin-0 and spin-1 nonets are included in the large $N_c$ ground state of QCD. Combined information from lattice data, dispersion relations and sum rules is being used \cite{Hohler2014,Hohler2016} to address the question of whether the axial vector $a_1\left(1260\right)$ meson achieves degeneracy with the $\rho\left(780\right)$ meson in the context of chiral symmetry restoration in relativistic heavy ion collisions. This might help to clarify the dispute regarding the nature of opposite parity states as being chiral partners or not.

The main goal of this paper is to generalize the result of previous works \cite{model1,model2,SQM} and include the spin-1 degrees of freedom in the effective meson Lagrangian together with corresponding important accompanying effects due to explicit chiral symmetry breaking. To fix the parameters of the model we calculate the masses of spin-0 and spin-1 low-lying meson states. The applications of the obtained model will be considered elsewhere.       

The present paper is organized as follows. In section \ref{s2} we briefly present the construction of the effective multi-quark vertices of the model, which had already been thoroughly discussed in \cite{model1,model2} for the spin-0 case. We then proceed to bosonize the effective Lagrangian in section \ref{s3} using a functional integral approach by introducing the physical boson fields as well as a set of auxiliary fields corresponding to quark bilinear structures. The auxiliary part of the functional integration is carried out in a stationary phase approximation (SPA) in subsection \ref{s3ss1}, and the Gaussian quark integration is performed  using a heat kernel technique in subsection \ref{s3ss2}. After the bosonization procedure, we focus on the quadratic part of the bosonized Lagrangian in section \ref{s4}, where we address the mixing between spin 0 and spin 1 boson fields and explain the necessary steps to get the meson kinetic and mass terms in standard form. Also in section \ref{s4}, the weak decay constants of the pseudoscalar mesons are computed from the quadratic part of the bosonized Lagrangian using the PCAC hypothesis. Finally, in section \ref{s5} we discuss the fitting of the model's parameters, with a particular focus on the possibility of reproducing the whole low-lying spin-0 and spin-1 meson spectra.

\section{Effective Multi-Quark Interactions 
\label{s2}}

Here we provide a brief review of the assumptions and procedure behind the construction of the effective meson Lagrangian. We refer to \cite{model1,model2} for a detailed description.  Then, we extend these ideas to the case with vector and axial-vector mesons.

The dynamical breaking of chiral symmetry in the light quark sector ($u,d$ and $s$ flavors) is proven to be a crucial mechanism for understanding the non-perturbative regime of QCD. An effective description of such regime \` a la Wilson requires a characteristic cut-off scale $\Lambda$ to be of the order of the spontaneous chiral symmetry breaking scale $\Lambda_{\chi SB}\sim 1\,\mbox{GeV}$, and presents itself as a natural expansion parameter for a chiral Lagrangian based on effective multi-quark vertices. The leading order effective Lagrangian includes local four-fermion couplings normalized to the $\Lambda$ cut-off scale. The higher-dimension multi-quark operators are responsible for next to leading order corrections in the description of the low-energy physics and correspondingly normalized to higher powers of $\Lambda$.

Furthermore, the explicit breaking of chiral symmetry due to finite current quark masses is extraneous to the strong interaction itself and may be realized by allowing the quarks to interact with an external source $\chi$; this approach facilitates the inclusion of the most general set of explicit symmetry breaking terms which are relevant at the order in $\Lambda$ and $N_c$ to which chirally symmetric terms are included.

Following the standard procedure, we define the quark bilinears (currents)
\begin{equation}
\label{currents}
s_a=\bar{q} \lambda_a q, \quad p_a = \bar{q} i \gamma_5 \lambda_a q, \quad
v_a^{\mu} = \bar{q} \gamma^{\mu} \lambda_a q, \quad
a_a^{\mu}  = \bar{q} \gamma^{\mu} \gamma_5 \lambda_a q,
\end{equation}
where $q$ is the quark field; $\gamma^{\mu}$ and $\gamma_5$ are Dirac matrices; the index $a$ takes on the values $0,1,\dots,8$; $\lambda_a$ are standard $U\left(3\right)$ matrices, where $\lambda_0 = \sqrt{\frac{2}{3}}\times\mathds{1}$ and the rest are the conventional $SU\left(3\right)$ Gell-Mann matrices, which obey the trace orthonormality condition $\mbox{tr}(\lambda_a\lambda_b)=2\delta_{ab}$. 

Using eq.(\ref{currents}) and completeness relations $\sum_{a=0}^8 (\lambda_a)_{ij}(\lambda_a)_{mn}=2\delta_{in}\delta_{jm}$, one may obtain the flavor components of the quark bilinears 
\begin{align}
\label{sb}
&\Sigma_{ij}=\frac{1}{2}\sum_{a=0}^8\left(s_a+ip_a\right)(\lambda_a)_{ij}=2\bar q_{Rj}q_{Li}, \\
\label{asb}
&\Sigma^\dagger_{ij}=\frac{1}{2}\sum_{a=0}^8\left(s_a-ip_a\right)(\lambda_a)_{ij}=2\bar q_{Lj}q_{Ri},
\end{align}
\begin{align}
\label{rv}
&R^{\mu}_{ij}=\frac{1}{2}\sum_{a=0}^8\left(v_a^{\mu}+a_a^{\mu}\right)(\lambda_a)_{ij}=2\bar q_{Rj}\gamma^\mu q_{Ri}, \\
\label{lv}
&L^{\mu}_{ij}  = \frac{1}{2}\sum_{a=0}^8 \left(v_a^{\mu} - a_a^{\mu}\right) (\lambda_a)_{ij}=2\bar q_{Lj}\gamma^\mu q_{Li}.
\end{align}
Here $q_R=P_Rq$, $q_L=P_Lq$, $\bar q_R=\bar q P_L$, $\bar q_L=\bar qP_R$, where $P_{R,L}=\frac{1}{2}(1\pm\gamma_5)$ are the right and left chiral projection operators. The action of the $U(3)_R\times U(3)_L$ group on quark fields is described by the unitary matrices $V_R$ and $V_L$: $q'(x)=(V_RP_R+V_LP_L)q(x)=V_Rq_R+V_Lq_L$. As a result, we find
\begin{equation}
\Sigma' = V_L \Sigma V_R^\dagger, \quad \Sigma^{'\dagger}=V_R\Sigma^\dagger V_L^\dagger, \quad R^{'\mu}=V_R R^{\mu} V_R^\dagger, \quad L^{'\mu}=V_L L^{\mu} V_L^\dagger.
\end{equation}

The terms of the effective multi-quark Lagrangian are built from the quark bilinears $\Sigma$, $\Sigma^\dagger$, $R^{\mu}$, $L^{\mu}$, the scale $\Lambda$, and the external source $\chi$ (the field $\chi$ is assumed to transform as $\Sigma$ and finally will be used to introduce explicit symmetry breaking effects), in a way which respects hermiticity, Lorentz and chiral invariance, as well as discrete symmetries such as parity and charge conjugation. Dimensional analysis together with the restriction to terms which contribute to the effective potential at $\Lambda \to \infty$ are employed in the selection of the effective terms which are considered relevant. These include the well-known four-, six- and eight-quark terms
\begin{align}
\label{main0}
\mathcal{L}_{int} & = \frac{\bar{G}}{\Lambda^2} \tr{\left(\Sigma^\dagger \Sigma\right)} + \frac{\bar{\kappa}}{\Lambda^5} \left(\det{\Sigma} + \det{\Sigma^\dagger}\right) \nonumber \\
& + \frac{\bar{g}_1}{\Lambda^8} \left(\tr{\Sigma^\dagger \Sigma}\right)^2 + \frac{\bar{g}_2}{\Lambda^8} \tr{\left(\Sigma^\dagger \Sigma \Sigma^\dagger \Sigma\right)},
\end{align}
among which the $U_A\left(1\right)$ breaking 't Hooft determinant (proportional to $\bar\kappa$) is included. Additionally, the 11 explicit symmetry breaking spin-0 terms are considered in \cite{model1,model2}
\begin{eqnarray}
\label{0esb}
\mathcal{L}_0 & = &-\tr{\left(\Sigma^\dagger \chi + \chi^\dagger \Sigma\right)}, \nonumber \\
\mathcal{L}_1 & = &- \frac{\bar{\kappa}_1}{\Lambda} \epsilon_{ijk} \epsilon_{mnl} \Sigma_{im} \chi_{jn} \chi_{kl} + h.c., \nonumber \\
\mathcal{L}_2 & = &\frac{\bar{\kappa}_2}{\Lambda^3} \epsilon_{ijk} \epsilon_{mnl} \Sigma_{im} \Sigma_{jn} \chi_{kl} + h.c., \nonumber \\
\mathcal{L}_3 & = &\frac{\bar{g}_3}{\Lambda^6} \tr{\left(\Sigma^\dagger \Sigma \Sigma^\dagger \chi\right)} + h.c., \nonumber \\
\mathcal{L}_4 & = &\frac{\bar{g}_4}{\Lambda^6} \tr{\left(\Sigma^\dagger \Sigma\right)} \tr{\left(\Sigma^\dagger \chi\right)} + h.c., \nonumber \\
\mathcal{L}_5 & = &\frac{\bar{g}_5}{\Lambda^4} \tr{\left(\Sigma^\dagger \chi \Sigma^\dagger \chi\right)} + h.c., \nonumber \\
\mathcal{L}_6 & = &\frac{\bar{g}_6}{\Lambda^4} \tr{\left(\Sigma^\dagger \Sigma \chi^\dagger \chi\right)} + h.c., \nonumber \\
\mathcal{L}_7 & = &\frac{\bar{g}_7}{\Lambda^4} \left(\tr{\Sigma^\dagger \chi} + h.c.\right)^2, \nonumber \\
\mathcal{L}_8 & = &\frac{\bar{g}_8}{\Lambda^4} \left(\tr{\Sigma^\dagger \chi} - h.c.\right)^2, \nonumber \\
\mathcal{L}_9 & = &\frac{\bar{g}_9}{\Lambda^2} \tr{\left(\Sigma^\dagger \chi \chi^\dagger \chi\right)} + h.c., \nonumber \\
\mathcal{L}_{10} & = & \frac{\bar{g}_{10}}{\Lambda^2} \tr{\left(\Sigma^\dagger \chi\right)} \tr{\left(\chi^\dagger \chi\right)} + h.c.
\end{eqnarray} 

In these expressions, the barred $G$, $g$'s and $\kappa$'s are dimensionless effective couplings; the traces and determinants refer to flavor space only, and $\epsilon_{ijk}$ is the Levi-Civita symbol in flavor space. Both $\mathcal{L}_{int}$ and the various $\mathcal{L}_i$ terms have already been discussed in \cite{model1,model2}, where it has been argued that they form a (spin-0) complete set in an expansion in $N_c$, with the term proportional to $\bar G$ and $\mathcal{L}_0$ being the leading order contributions ($\mathcal O\left(N_c^1\right)$) and the other terms being the higher order in $1/N_c$ expansion. This classification of multi-quark effective terms within an $N_c$ expansion has been shown to be consistent with the expansion in $1/\Lambda$, i.e. the full Lagrangian thus considered consists of all the terms which yield a contribution to the effective potential up to the order $\mathcal O(\Lambda^0)$. The $N_c$ counting assignments for the effective couplings which follow thereof have been pointed out as $G,\kappa_1,g_9,g_{10} \sim N_c^{-1}$, $\kappa_2,g_5,g_6,g_7,g_8 \sim N_c^{-2}$, $\kappa,g_3,g_4 \sim N_c^{-3}$, and $g_1,g_2 \sim N_c^{-4}$, with $\Lambda \sim N_c^0$. Furthermore, it has been pointed out that the terms proportional to $\kappa,\kappa_1,\kappa_2,g_1,g_4,g_7,g_8,g_{10}$ trace OZI rule violating affects, while those proportional to $g_2,g_3,g_5,g_6,g_9$ express an admixture of four-quark components $\bar{q}q\bar{q}q$ to the $\bar{q}q$ one. As a final remark, it has been noted that the terms proportional to $\kappa_1,g_9,g_{10}$, which are bilinear in quark fields, may be related to the known Kaplan-Manohar ambiguity \cite{Kaplan86} in the definition of current quark masses, so that these couplings may be set to 0 without loss of generality.

In this work, we do not consider multi-quark effective terms with derivatives. In a local multi-quark Lagrangian, derivative interactions contribute only (through bosonization) to radial excitations of the meson fields \cite{Volkov97,Andrianov2005}. For modelling the low-lying states, these contributions are then dispensable. Derivative terms would further allow for non-homogeneous quark condensates, a feature which is also beyond the scope of the present work.

To extend the above ideas to spin-1 states we follow here the same logic. As a result, we were able to identify 13 new terms which include $R^{\mu}$ and $L^{\mu}$ quark bilinears:
\begin{eqnarray*}
\mathcal{L}'_1 & = &\frac{\bar{w}_1}{\Lambda^2} \tr{\left(R^{\mu}R_{\mu} + L^{\mu}L_{\mu}\right)}, \nonumber \\
\mathcal{L}'_2 & = &\frac{\bar{w}_2}{\Lambda^8} \left[\tr{\left(R^{\mu}R_{\mu} + L^{\mu}L_{\mu}\right)}\right]^2, \nonumber \\
\mathcal{L}'_3 & = &\frac{\bar{w}_3}{\Lambda^8} \left[\tr{\left(R^{\mu}R_{\mu} - L^{\mu}L_{\mu}\right)}\right]^2, \nonumber \\
\mathcal{L}'_4 & = &\frac{\bar{w}_4}{\Lambda^8} \tr{\left(R^{\mu}R^{\nu}R_{\mu}R_{\nu} + L^{\mu}L^{\nu}L_{\mu}L_{\nu}\right)}, \nonumber \\
\mathcal{L}'_5 & = &\frac{\bar{w}_5}{\Lambda^8} \tr{\left(R^{\mu}R_{\mu}R^{\nu}R_{\nu} + L^{\mu}L_{\mu}L^{\nu}L_{\nu}\right)}, \nonumber \\
\mathcal{L}'_6 & = &\frac{\bar{w}_6}{\Lambda^8} \tr{\left(R^{\mu}R_{\mu} + L^{\mu}L_{\mu}\right)} \tr{\left(\Sigma^\dagger \Sigma\right)}, \nonumber \\
\mathcal{L}'_7 & = &\frac{\bar{w}_7}{\Lambda^8} \tr{\left(\Sigma^\dagger L^{\mu} \Sigma R_{\mu}\right)}, \nonumber \\
\mathcal{L}'_8 & = &\frac{\bar{w}_8}{\Lambda^8} \tr{\left(\Sigma^\dagger \Sigma R^{\mu} R_{\mu} + \Sigma \Sigma^\dagger L^{\mu} L_{\mu}\right)}, \nonumber \\
\mathcal{L}'_9 & = &\frac{\bar{w}_9}{\Lambda^6} \tr{\left(R^{\mu}R_{\mu} + L^{\mu}L_{\mu}\right)} \tr{\left(\Sigma^\dagger \chi + \Sigma \chi^\dagger\right)}, \nonumber \\
\mathcal{L}'_{10} & = &\frac{\bar{w}_{10}}{\Lambda^6} \tr{\left(\chi^\dagger L^{\mu} \Sigma R_{\mu} + \Sigma^\dagger L^{\mu} \chi R_{\mu}\right)}, \nonumber \\
\end{eqnarray*}
\begin{eqnarray}
\label{main1}
\mathcal{L}'_{11} & = &\frac{\bar{w}_{11}}{\Lambda^6} \tr{\left[\left(\Sigma^\dagger \chi + \chi^\dagger \Sigma\right) R^{\mu} R_{\mu} + \left(\Sigma \chi^\dagger + \chi \Sigma^\dagger\right) L^{\mu} L_{\mu} \right]}, \nonumber \\
\mathcal{L}'_{12} & = &\frac{\bar{w}_{12}}{\Lambda^4} \tr{\left(\chi^\dagger L^{\mu} \chi R_{\mu}\right)}, \nonumber \\
\mathcal{L}'_{13} & = &\frac{\bar{w}_{13}}{\Lambda^4} \tr{\left(\chi^\dagger \chi R^{\mu} R_{\mu} + \chi \chi^\dagger L^{\mu} L_{\mu}\right)}.
\end{eqnarray}
Also here, the $\bar w$'s are dimensionless effective couplings. $\mathcal{L}'_1$ is the only four-quark term and it is the spin-1 analogue of the term proportional to $G_V$ in the usual NJL-type model. Terms $\mathcal{L}'_2$ to $\mathcal{L}'_5$ are purely spin-1 eight-quark terms, while $\mathcal{L}'_6$ to $\mathcal{L}'_8$ represent mixed eight-quark terms involving both spin-0  and spin-1 components. Finally, the five terms $\mathcal{L}'_9$ to $\mathcal{L}'_{13}$ complete the set of explicit symmetry breaking terms which are relevant at next-to-leading order. The terms proportional to $w_2,w_3,w_6,w_9$ express OZI rule violating effects, while those proportional to $w_4,w_5,w_7,w_8,w_{10},w_{11},w_{12},w_{13}$ are related with an admixture of four-quark $\bar{q}q\bar{q}q$ components to the $\bar{q}q$ one.

\section{Functional Bosonization 
\label{s3}}

In order to have an effective model in terms of hadronic degrees of freedom, we proceed to bosonize the multi-quark Lagrangian. The starting point is the functional integral
\begin{equation}
\label{Z}
Z = \int\! \mathcal{D}q \mathcal{D}\bar{q}\, e^{\,i\!\int\! d^4 x \mathcal{L}(x)},
\end{equation}
where the Lagrangian density $\mathcal L(x)$ is given by
\begin{equation}
\mathcal{L} = i\bar{q} \gamma^{\mu} \partial_{\mu} q + \mathcal{L}_{int} + \sum_{i=0}^{10} \mathcal{L}_i + \sum_{i=1}^{15} \mathcal{L}'_i.
\end{equation}

Next, we use the functional representation of unity \cite{Reinhardt88}
\begin{align}
1 & = \int \mathcal{D} s_a \mathcal{D} p_a \mathcal{D} v^{\mu}_a \mathcal{D} a^{\mu}_a \delta\left(s_a - \bar{q} \lambda_a q\right) \delta\left(p_a - i\bar{q} \lambda_a \gamma_5 q\right)\delta\left(v^{\mu}_a - \bar{q} \lambda_a \gamma^{\mu} q\right) \nonumber \\
& \times\delta\left(a^{\mu}_a - \bar{q} \lambda_a \gamma^{\mu} \gamma_5 q\right)= \int \mathcal{D} \sigma_a \mathcal{D} \phi_a \mathcal{D} V_{a\mu} \mathcal{D} A_{a\mu} \int \mathcal{D} s_a \mathcal{D} p_a \mathcal{D} v^{\mu}_a \mathcal{D} a^{\mu}_a \nonumber \\
& \times e^{i\int d^4 x \left[\sigma_a\left(s_a - \bar{q} \lambda_a q\right) + \phi_a\left(p_a - i\bar{q} \lambda_a \gamma_5 q\right) + V_{a\mu}\left(v^{\mu}_a - \bar{q} \lambda_a \gamma^{\mu} q\right) + A_{a\mu}\left(a^{\mu}_a - \bar{q} \lambda_a \gamma^{\mu} \gamma_5 q\right)\right]}
\end{align}
as a tool to introduce into (\ref{Z}) the auxiliary bosonic fields $\sigma = \sigma_a \lambda_a$, $\phi = \phi_a \lambda_a$, $V^{\mu} = V^{\mu}_a \lambda_a$ and $A^{\mu} = A^{\mu}_a \lambda_a$. The resulting functional integral reads
\begin{align}
\label{Z2}
Z & = \int \mathcal{D} \sigma_a \mathcal{D} \phi_a \mathcal{D} V_{a\mu} \mathcal{D} A_{a\mu} \int \mathcal{D} s_a \mathcal{D} p_a \mathcal{D} v^{\mu}_a \mathcal{D} a^{\mu}_a \,e^{\,i\!\int\! d^4 x \mathcal{L}_{aux}} \nonumber \\
& \times \int \mathcal{D}q \mathcal{D}\bar{q}\, e^{\,i\! \int\! d^4 x \bar{q} \left(i\gamma^{\mu}\partial_{\mu} - \sigma - i\gamma_5\phi - \gamma^{\mu}V_{\mu} - \gamma^{\mu}\gamma_5A_{\mu}\right)q},
\end{align}
where we have defined the Lagrangian density 
\begin{equation}
\mathcal{L}_{aux}\left(s,p,v,a\right)=\mathcal{L}_{int}+\sum_{i=2}^{8}\mathcal{L}_i +\sum_{i=1}^{13}\mathcal{L}'_i+s_a\left(\sigma_a -m_a\right)+p_a\phi_a + v_a^{\mu}V_{a\mu} + a_a^{\mu}A_{a\mu}
\end{equation}
with the previously defined quark bilinears rewritten as functions of the auxiliary boson fields $s_a, p_a, v_a^{\mu}, a_a^{\mu}$ ($\kappa_1 = g_9 = g_{10} = 0$), and with the substitution $\chi = m/2$, $m$ being the current quarks mass matrix. These expressions read
\begin{align}
\mathcal{L}_{int} & = \frac{\bar{G}}{2\Lambda^2} \left(s_a^2 + p_a^2\right) + \frac{\bar{\kappa}}{4\Lambda^5} A_{abc}s_a\left(s_b s_c - 3 p_b p_c\right) + \frac{\bar{g}_1}{4\Lambda^8} \left(s_a^2 + p_a^2\right)^2 \nonumber \\
& + \frac{\bar{g}_2}{8\Lambda^8}\left[d_{abe}d_{cde}\left(s_a s_b + p_a p_b\right)\left(s_c s_d + p_c p_d\right) + 4f_{abe}f_{cde} s_a s_c p_b p_d\right],
\end{align}
and
\begin{align}
\mathcal{L}_2 & = \frac{3\bar{\kappa}_2}{2\Lambda^3} A_{abc} m_a \left(s_b s_c - p_b p_c\right), \nonumber \\
\mathcal{L}_3 & = \frac{\bar{g}_3}{4\Lambda^6} m_a \left[d_{abe}d_{cde} s_b \left(s_c s_d + p_c p_d\right) - 2 f_{abe}f_{cde} p_b p_c s_d\right], \nonumber \\
\mathcal{L}_4 & = \frac{\bar{g}_4}{2\Lambda^6} \left(s_a^2 + p_a^2\right) s_b m_b, \nonumber \\
\mathcal{L}_5 & = \frac{\bar{g}_5}{4\Lambda^4} \left(d_{abe}d_{cde} - f_{abe}f_{cde}\right) m_b m_d \left(s_a s_c - p_a p_c\right), \nonumber \\
\mathcal{L}_6 & = \frac{\bar{g}_6}{4\Lambda^4} d_{abe}d_{cde} m_a m_b \left(s_c s_d - p_c p_d\right), \nonumber \\
\mathcal{L}_7 & = \frac{\bar{g}_7}{\Lambda^4} \left(s_a m_a\right)^2, \nonumber \\
\mathcal{L}_8 & = \frac{\bar{g}_8}{\Lambda^4} \left(p_a m_a\right)^2,
\end{align}
and correspondingly for spin-1 fields
\begin{align*}
\mathcal{L}'_1 & = \frac{\bar{w}_1}{\Lambda^2} \left(v^{\mu}_a v_{a\mu} + a^{\mu}_a a_{a\mu}\right), \nonumber \\
\mathcal{L}'_2 & = \frac{\bar{w}_2}{\Lambda^8} \left(v^{\mu}_a v_{a\mu} + a^{\mu}_a a_{a\mu}\right)^2, \nonumber \\
\mathcal{L}'_3 & = \frac{4\bar{w}_3}{\Lambda^8} \left(v^{\mu}_a a_{a\mu}\right)^2, \nonumber \\
\mathcal{L}'_4 & = \frac{\bar{w}_4}{4\Lambda^8} \left(d_{ace}d_{bde} - f_{ace}f_{bde}\right) \left[\left(v^{\mu}_a v_{b\mu} + a^{\mu}_a a_{b\mu}\right) \left(v^{\nu}_c v_{d\nu} + a^{\nu}_c a_{d\nu}\right) \right. \nonumber \\
& \left. + \left(v^{\mu}_a a_{b\mu} + a^{\mu}_a v_{b\mu} \right) \left(v^{\nu}_c a_{d\nu} + a^{\nu}_c v_{d\nu} \right) \right], \nonumber \\
\mathcal{L}'_5 & = \frac{\bar{w}_5}{4\Lambda^8} d_{abe}d_{cde} \left[\left(v^{\mu}_a v_{b\mu} + a^{\mu}_a a_{b\mu}\right) \left(v^{\nu}_c v_{d\nu} + a^{\nu}_c a_{d\nu}\right) + \left(v^{\mu}_a a_{b\mu} + a^{\mu}_a v_{b\mu} \right) \left(v^{\nu}_c a_{d\nu} + a^{\nu}_c v_{d\nu} \right) \right], \nonumber \\
\mathcal{L}'_6 & = \frac{\bar{w}_6}{2\Lambda^8} \left(v^{\mu}_a v_{a\mu} + a^{\mu}_a a_{a\mu}\right) \left(s_b^2 + p_b^2\right), \nonumber \\
\end{align*}
\begin{align}
\mathcal{L}'_7 & = \frac{\bar{w}_7}{8\Lambda^8} \left[ \left(d_{ace}d_{bde} - f_{ace}f_{bde}\right) \left(s_a s_b + p_a p_b\right) \left(v^{\mu}_c v_{d\mu} - a^{\mu}_c a_{d\mu}\right) \right.  \nonumber \\
& \left. + \left(f_{ace}d_{bde} + d_{ace}f_{bde}\right) \left(s_a p_b - p_a s_b\right) \left(a^{\mu}_c v_{d\mu} - v^{\mu}_c a_{d\mu}\right) \right], \nonumber \\
\mathcal{L}'_8 & = \frac{\bar{w}_8}{4\Lambda^8}\, d_{cde} \left[d_{abe} \left(s_a s_b + p_a p_b\right) \left(v^{\mu}_c v_{d\mu} + a^{\mu}_c a_{d\mu}\right) \right.  \nonumber \\
& \left. + f_{abe} \left(p_a s_b - s_a p_b\right) \left(v^{\mu}_c a_{d\mu} + a^{\mu}_c v_{d\mu}\right) \right], \nonumber \\
\mathcal{L}'_9 & = \frac{\bar{w}_9}{\Lambda^6} \left(v^{\mu}_a v_{a\mu} + a^{\mu}_a a_{a\mu}\right) s_b m_b, \nonumber \\
\mathcal{L}'_{10} & = \frac{\bar{w}_{10}}{8\Lambda^6} \left[ \left(d_{ace}d_{bde} - f_{ace}f_{bde}\right) \left(s_a m_b + s_b m_a\right) \left(v^{\mu}_c v_{d\mu} - a^{\mu}_c a_{d\mu}\right) \right.  \nonumber \\
& \left. + \left(f_{ace}d_{bde} + d_{ace}f_{bde}\right) \left(p_a m_b - p_b m_a\right) \left(v^{\mu}_c a_{d\mu} - a^{\mu}_c v_{d\mu}\right) \right], \nonumber \\
\mathcal{L}'_{11} & = \frac{\bar{w}_{11}}{2\Lambda^6} m_a \left[ d_{abe} s_b \left(v^{\mu}_c v_{d\mu} + a^{\mu}_c a_{d\mu}\right) - f_{abe} p_b \left(v^{\mu}_c a_{d\mu} + a^{\mu}_c v_{d\mu}\right) \right] d_{cde},  \nonumber \\
\mathcal{L}'_{12} & = \frac{\bar{w}_{12}}{8\Lambda^4} \left(d_{ace}d_{bde} - f_{ace}f_{bde}\right)m_a m_b \left(v^{\mu}_c v_{d\mu} - a^{\mu}_c a_{d\mu}\right), \nonumber \\
\mathcal{L}'_{13} & = \frac{\bar{w}_{13}}{4\Lambda^4} d_{abe}d_{cde} m_a m_b \left(v^{\mu}_c v_{d\mu} + a^{\mu}_c a_{d\mu}\right).
\end{align}
In these expressions, $f_{abc}$ are the antisymmetric structure constants of a Lie algebra ($\left[\lambda_a,\lambda_b\right] = 2i f_{abc} \lambda_c$) related to the $U\left(3\right)$ flavor group, while $d_{abc}$ are the corresponding symmetric constants ($\left\lbrace\lambda_a,\lambda_b\right\rbrace = 2 d_{abc} \lambda_c$), and
\begin{equation}
A_{abc} = \frac{1}{3!} \epsilon_{ijk} \epsilon_{mnl} \left(\lambda_a\right)_{im} \left(\lambda_b\right)_{jn} \left(\lambda_c\right)_{kl}
\end{equation}
is a totally symmetric tensor in flavor linear space.

In the Nambu-Goldstone realization of chiral symmetry, the scalar field $\sigma$ develops a finite vacuum expectation value $\langle\sigma\rangle = M$. In order to properly describe excitations around the true unsymmetric vacuum we make a shift $\sigma \to \sigma + M$ in (\ref{Z2}). $M$ may be interpreted as a constituent quark mass matrix. Defining $\Delta = M - m$, we may rewrite (\ref{Z2}) as
\begin{align}
\label{Z3}
Z & = \int \mathcal{D} \sigma_a \mathcal{D} \phi_a \mathcal{D} V_{a\mu} \mathcal{D} A_{a\mu} \int \mathcal{D} s_a \mathcal{D} p_a \mathcal{D} v^{\mu}_a \mathcal{D} a^{\mu}_a\, e^{\,i\!\int\! d^4 x \mathcal{L}_{aux}} \nonumber \\
& \times \int \mathcal{D}q \mathcal{D}\bar{q}\, e^{\,i\!\int\! d^4 x \bar{q} \left(i\gamma^{\mu}\partial_{\mu} - M - \sigma - i\gamma_5\phi - \gamma^{\mu}V_{\mu} - \gamma^{\mu}\gamma_5A_{\mu}\right)q} \nonumber \\
& = \int \mathcal{D} \sigma_a \mathcal{D} \phi_a \mathcal{D} V_{a\mu} \mathcal{D} A_{a\mu}\, e^{\,i\!\int\! d^4 x \left(\mathcal{L}_{SPA} + \mathcal{L}_{HK}\right)}.
\end{align}
Here the auxiliary Lagrangian density ${\mathcal L}_{aux}$ is given now by 
\begin{equation}
{\mathcal L}_{aux}=\mathcal{L}_{int} + \sum_{i=2}^{8} \mathcal{L}_i + \sum_{i=1}^{13} \mathcal{L}'_i + s_a\left(\sigma_a - \Delta_a\right) + p_a\phi_a + v_a^{\mu}V_{a\mu} + a_a^{\mu}A_{a\mu}.
\end{equation}
The full bosonized Lagrangian appearing in the last line of (\ref{Z3}) comprises contributions from the integration over auxiliary fields,
\begin{equation}
\label{SPA-int}
e^{\,i\!\int\! d^4x\mathcal{L}_{SPA}} = \int \mathcal{D} s_a \mathcal{D} p_a \mathcal{D} v^{\mu}_a \mathcal{D} a^{\mu}_a \,e^{\,i\!\int\! d^4x\mathcal{L}_{aux}},
\end{equation}
and from the quark Gaussian integral
\begin{equation}
\label{HK-int}
e^{\,i\!\int\! d^4x\mathcal{L}_{HK}} = \int\mathcal{D}q \mathcal{D}\bar{q}\,e^{\,i\!\int\! d^4x\bar{q}\left(i\gamma^{\mu}\partial_{\mu} - M - \sigma - i\gamma_5\phi - \gamma^{\mu}V_{\mu} - \gamma^{\mu}\gamma_5A_{\mu}\right)q}.
\end{equation}
The former is performed using a stationary phase approximation (SPA), while the latter is computed with a modified heat kernel technique.

\subsection{Stationary Phase Aproximation 
\label{s3ss1}}

As was done in \cite{model1,model2}, the functional integration in (\ref{SPA-int}) is performed by means of a SPA; the auxiliary fields $s_a, p_a, v_a^\mu, a_a^\mu$ have no kinetic terms and yield the simple classical equations of motion
\begin{equation}
\label{SPA-conds}
\left.\frac{\partial \mathcal{L}_{aux}}{\partial s_a}\right|_{s_a = s_a^{st}} = \left.\frac{\partial \mathcal{L}_{aux}}{\partial p_a}\right|_{p_a = p_a^{st}} = \left.\frac{\partial \mathcal{L}_{aux}}{\partial v^{\mu}_a}\right|_{v^{\mu}_a = v^{\mu\, st}_a} = \left.\frac{\partial \mathcal{L}_{aux}}{\partial a^{\mu}_a}\right|_{a^{\mu}_a = a^{\mu\, st}_a} = 0.
\end{equation}

We seek solutions in the form of a series in powers of boson fields $\sigma, \phi, V^\mu_a, A^\mu_a$
\begin{align}
\label{SPA-exp}
s_a^{st} & = h_a + h_{ab}^{\left(1\right)} \sigma_b + h_{abc}^{\left(1\right)} \sigma_b \sigma_c + h_{abc}^{\left(2\right)} \phi_b \phi_c + H_{abc}^{\left(1\right)} V^{\mu}_b V_{c\mu} + H_{abc}^{\left(2\right)} A^{\mu}_b A_{c\mu} + \dots \nonumber \\
p_a^{st} & = h_{ab}^{\left(2\right)} \phi_b + h_{abc}^{\left(3\right)} \phi_b \sigma_c + H_{abc}^{\left(3\right)} V^{\mu}_b A_{c\mu} + \dots \nonumber \\
v^{\mu\, st}_a & = H_{ab}^{\left(1\right)} V^{\mu}_b + H_{abc}^{\left(4\right)} \sigma_b V^{\mu}_c + H_{abc}^{\left(5\right)} \phi_b A^{\mu}_c + \dots \nonumber \\
a^{\mu\, st}_a & = H_{ab}^{\left(2\right)} A^{\mu}_b + H_{abc}^{\left(6\right)} \phi_b V^{\mu}_c + H_{abc}^{\left(7\right)} \sigma_b A^{\mu}_c + \dots
\end{align}
By equating the coefficient of each monomial combination of fields in (\ref{SPA-conds}) to zero, we are able to express the several coefficients $h$ and $H$ appearing in (\ref{SPA-exp}) recursively in terms of the model parameters. The first such expression stems from the $\mathcal O\left(1\right)$ term in $\frac{\partial \mathcal{L}_{aux}}{\partial s_a}$ and yields an implicit cubic expression for the $h_a$. It turns out that $h_a = 0$ for $a \neq 0,3,8$ (i.e. only the diagonal components of $h= h_a \lambda_a$ are nonzero). We may choose to transform the index $a$ into a fundamental flavor basis $i=u,d,s$ with $h_a = e_{ai} h_i$ ($a=0,3,8$) and
\begin{equation}
e_{ai} = \frac{\left(\lambda_a\right)_{ii}}{2}=\frac{1}{2\sqrt{3}}\left(
\begin{array}{ccc}
\sqrt{2} & \sqrt{2} & \sqrt{2} \\
\sqrt{3} & -\sqrt{3} & 0 \\
1 & 1 & -2
\end{array} \right), \quad a=0,3,8.
\end{equation}
Then the system of three equations to find $h_i$ is 
\begin{align}
\label{hi-conds}
\Delta_i + \frac{h_i}{4} \left(4G + 2g_1 h^2 + 2g_4 mh\right) + \frac{g_2}{2} h_i^3 + \frac{\kappa}{4} t_{ijk} h_j h_k + \kappa_2 t_{ijk} h_j m_k \nonumber \\
+ \frac{m_i}{4} \left[ 3 g_3 h_i^2 + g_4 h^2 + 2\left(g_5 + g_6 \right) m_i h_i + 4 g_7 mh \right] = 0.
\end{align}
Here we use the definitions $h^2 = h_u^2 + h_d^2 + h_s^2$ and $mh = m_u h_u + m_d h_d + m_s h_s$. The $h_i$ are in direct connection with the quark condensates $\left\langle \bar{q}_i q_i \right\rangle$ which play the role of order parameters in the transition between Wigner-Weyl and Nambu-Goldstone realizations of chiral symmetry. The conditions (\ref{hi-conds}) had already been found in \cite{model1,model2}, a fact which indicates that the inclusion of vector modes in the model has no direct impact in the SPA conditions for the quark condensates.

Expressions for the two-index $h_{ab}$ and $H_{ab}$ coefficients may be computed from linear monomials in (\ref{SPA-conds}). The result is
\begin{align}
-2\left[h_{ab}^{\left(1\right)}\right]^{-1} & = \left(2G + g_1 h^2 + g_4 mh\right) \delta_{ab} + 4g_1 h_a h_b + 3 A_{abc} \left(\kappa h_c + 2\kappa_2 m_c\right) \nonumber \\
& + g_2 \left(d_{abe}d_{cde} + 2d_{ace}d_{bde}\right) h_c h_d+ g_3 \left(d_{abe}d_{cde} + d_{ace}d_{bde} + d_{ade}d_{bce}\right) h_c m_d \nonumber \\
& + 2g_4 \left(h_a m_b + h_b m_a\right) + g_5\left(d_{ace}d_{bde} - f_{ace}f_{bde}\right) m_c m_d + g_6 d_{abe}d_{cde} m_c m_d \nonumber \\
& + 4 g_7 m_a m_b,
\end{align}
\begin{align}
-2\left[h_{ab}^{\left(2\right)}\right]^{-1} & = \left(2G + g_1 h^2 + g_4 mh\right) \delta_{ab} - 3 A_{abc} \left(\kappa h_c + 2\kappa_2 m_c\right) + g_2 \left(d_{abe}d_{cde} + 2f_{ace}f_{bde}\right) h_c h_d \nonumber \\
& + g_3 \left(d_{abe}d_{cde} + f_{ace}f_{bde} + f_{ade}f_{bce}\right) h_c m_d - g_5\left(d_{ace}d_{bde} - f_{ace}f_{bde}\right) m_c m_d \nonumber \\ 
& + g_6 d_{abe}d_{cde} m_c m_d - 4 g_8 m_a m_b,
\end{align}
where again the inclusion of vector modes has no direct consequence on the above formulae, which have been previously obtained. 

On the opposite, two other coefficients $H_{ab}^{(1)}$ and $H_{ab}^{(2)}$ associated with vector and axial-vector terms are new. They are given by the following expressions 
\begin{align}
\label{Hab1}
-2\left[H_{ab}^{\left(1\right)}\right]^{-1} & = \left(4w_1 + w_6 h^2 + 2w_9 mh\right) \delta_{ab} + \frac{w_7}{2} \left(d_{ace}d_{bde} - f_{ace}f_{bde}\right) h_c h_d \nonumber \\
& + w_8 d_{abe}d_{cde} h_c h_d + w_{10} \left(d_{ace}d_{bde} - f_{ace}f_{bde}\right) h_c m_d + 2 w_{11} d_{abe}d_{cde} h_c m_d \nonumber \\
& + \frac{w_{12}}{2} \left(d_{ace}d_{bde} - f_{ace}f_{bde}\right) m_c m_d + w_{13} d_{abe}d_{cde} m_c m_d,
\end{align}
\begin{align}
\label{Hab2}
-2\left[H_{ab}^{\left(2\right)}\right]^{-1} & = \left(4w_1 + w_6 h^2 + 2w_9 mh\right) \delta_{ab} - \frac{w_7}{2} \left(d_{ace}d_{bde} - f_{ace}f_{bde}\right) h_c h_d \nonumber \\
& + w_8 d_{abe}d_{cde} h_c h_d - w_{10} \left(d_{ace}d_{bde} - f_{ace}f_{bde}\right) h_c m_d + 2 w_{11} d_{abe}d_{cde} h_c m_d \nonumber \\
& - \frac{w_{12}}{2} \left(d_{ace}d_{bde} - f_{ace}f_{bde}\right) m_c m_d + w_{13} d_{abe}d_{cde} m_c m_d.
\end{align}
\noindent The striking similarity between the two expressions (\ref{Hab1}) and (\ref{Hab2}), where the only difference is in the signs of those terms proportional to $w_7$, $w_{10}$ and $w_{12}$, is a noteworthy aspect with important consequences regarding the fitting of the model's parameters. Another interesting feature of these expressions is the apparent decoupling between spin-0 and spin-1 related parameters; the new $w$'s enter only in the new coefficients $H_{ab}$, with none of the old parameters entering alongside.

The three-index coefficients $h_{abc}$ and $H_{abc}$ are determined from quadratic (or bilinear) monomials appearing in (\ref{SPA-conds}), which we collect for future reference in Appendix \ref{app1}. We remark that the apparent disconnection between spin-0 and spin-1 related parameters is again manifest in these expressions.

This procedure can be extended to obtain the higher index coefficients. As a result the coefficients $h_a$ (and couplings of multi-quark interactions) fully determine all of them. Finally, all these recursion relations may be used to find the contribution to the bosonized Lagrangian density resulting from the SPA functional integration; it reads (up to cubic terms in the fields)
\begin{align}
\label{SPA-Lag}
\mathcal{L}_{SPA} \left(\sigma, \phi, V_{\mu}, A_{\mu}\right)=h_a \sigma_a &+ \frac{1}{2} \left(h_{ab}^{\left(1\right)} \sigma_a \sigma_b + h_{ab}^{\left(2\right)} \phi_a \phi_b + H_{ab}^{\left(1\right)} V^{\mu}_a V_{b\mu} + H_{ab}^{\left(2\right)} A^{\mu}_a A_{b\mu} \right) \nonumber \\
& + \sigma_a \left(\frac{1}{3} h_{abc}^{\left(1\right)} \sigma_b \sigma_c + h_{abc}^{\left(2\right)} \phi_b \phi_c + H_{abc}^{\left(1\right)} V^{\mu}_b V_{c\mu} + H_{abc}^{\left(2\right)} A^{\mu}_b A_{c\mu}\right) \nonumber \\
& + H_{abc}^{\left(3\right)} \phi_a V^{\mu}_b A_{c\mu} + \ldots
\end{align}
Equation (\ref{SPA-Lag}) sheds light onto the physical role played by the various $h$ and $H$ coefficients. The $h_a$ are related with the amplitude of the tadpole $\sigma_a$ terms, i.e. with the vacuum expectation value of the $\sigma_a$ field. The two-index coefficients express SPA-contributions to the masses of the boson fields, while the three-index coefficients yield contributions to the couplings of effective three-field interaction vertices.

\subsection{Quark Determinant 
\label{s3ss2}}

The calculation of the quark determinant contribution to the bosonized Lagrangian is performed with a generalized heat kernel technique \cite{HK1,HK2,HK3} which accommodates the possibility of a non-degenerate mass matrix $M$. The method consists of a suitable resummation of the heat kernel series which ensures that, to each order in the modified series expansion, the resulting contribution to the bosonized Lagrangian remains consistent with the pre-determined chiral symmetry requirements. The Gaussian functional integral (up to an overall unessential constant) may be rewritten in an Euclidean metric as
\begin{equation}
\int \mathcal{D} q \mathcal{D} \bar{q} \exp\left(-\int\! d^4 x_E\, \bar{q} D_E q\right) = \det D_E,
\end{equation}
with the Dirac operator 
\begin{equation}
D_E = i \gamma_{\alpha}\partial_{\alpha} - M - \sigma - i\gamma_5\phi + \gamma_{\alpha}V_{\alpha} + \gamma_{\alpha}\gamma_5 A_{\alpha}.
\end{equation}

The contribution of the chiral determinant to the real part of the effective action can be found in accord with the following formal manipulations
\begin{equation}
\det D_E\to\det \left| D_E \right| = \det\sqrt{D_E^\dagger D_E}=e^{\ln\det\sqrt{D_E^\dagger D_E}}=e^{\frac{1}{2}\mbox{\scriptsize tr}\ln D_E^\dagger D_E},
\end{equation}
by which the modified heat kernel expansion results in a series of the form
\begin{equation}
\frac{1}{2}\mbox{tr}\ln D_E^\dagger D_E = - \int \frac{d^4 x_E}{32 \pi^2} \sum_{n=0}^{\infty} I_{n-1} \tr{\left(b_n\right)}.
\end{equation}
Here
\begin{equation}
I_n = \frac{1}{3} \sum_{i=u,d,s} J_n \left(M_i^2\right) 
\end{equation}
and
\begin{equation}
J_n \left(M_i^2\right) = \int_0^{\infty} \frac{d\tau}{\tau^{2-n}} \rho\left(\tau\Lambda^2\right) e^{-\tau M_i^2}
\end{equation}
are the Schwinger's proper-time integrals in which a regulating kernel $\rho\left(\tau\Lambda^2\right)$ is specified as a Pauli-Villars type regulator with double subtractions \cite{Pauli-Villars,regul}
\begin{equation}
\rho\left(\tau\Lambda^2\right) = 1 - \left(1+\tau\Lambda^2\right) e^{-\tau\Lambda^2}.
\end{equation}

The generalized Seeley-DeWitt coefficients $b_n$ for the spin-0 version of the model have been obtained in \cite{model1,model2}. These can be translated into the appropriate form for the model under study through the procedure described in \cite{Salcedo}. The first three coefficients read
\begin{align}
b_0 & = \mathds{1} \\
b_1 & = -Y \\
b_2 & = \frac{Y^2}{2} + \frac{\Delta_{12}}{2} \lambda_3 Y + \frac{\Delta_{13}+\Delta_{23}}{2\sqrt{3}} \lambda_8 Y - \frac{\Gamma_{\alpha\beta}^2}{12},
\end{align}
where $\Delta_{ij} = M_i^2 - M_j^2$ and with the following definitions:
\begin{equation}
Y =\sigma^2+\left\lbrace\sigma, M\right\rbrace +\phi^2-i\gamma_5\left[\phi,\sigma + M\right]+i\gamma_{\alpha}\nabla'_{\alpha}\left(\sigma+i\gamma_5\phi\right)- \frac{i}{4} \left[\gamma_{\alpha},\gamma_{\beta}\right] \Gamma_{\alpha\beta},   
\end{equation}
\begin{align}
\nabla'_{\alpha}\sigma & = \partial_{\alpha} \sigma - \left\{ A_{\alpha},\phi \right\} - i \left[V_{\alpha}, \sigma + M\right] \\
\nabla'_{\alpha} \phi & = \partial_{\alpha} \phi + \left\{ A_{\alpha},\sigma + M \right\} - i \left[V_{\alpha}, \phi\right], \\
V_{\alpha\beta} & = \partial_{\alpha} V_{\beta} - \partial_{\beta} V_{\alpha} - i \left\lbrack V_{\alpha}, V_{\beta} \right\rbrack - i \left\lbrack A_{\alpha}, A_{\beta} \right\rbrack \\
A_{\alpha\beta} & = \partial_{\alpha} A_{\beta} - \partial_{\beta} A_{\alpha} - i \left\lbrack V_{\alpha}, A_{\beta} \right\rbrack - i \left\lbrack A_{\alpha}, V_{\beta} \right\rbrack, \\
\Gamma_{\alpha\beta} &= V_{\alpha\beta} + \gamma_5 A_{\alpha\beta}.
\end{align}

Just as in the spin-0 version of the model, both $b_1$ and $b_2$ provide contributions to the spin-0 boson masses and to the gap equations. The latter arise from the requirement that the $\sigma$ tadpole term of the overall bosonized Lagrangian $\mathcal{L}_{bos} = \mathcal{L}_{SPA} + \mathcal{L}_{HK}$ should vanish. To order $n=2$ in the heat kernel expansion the gap equations are
\begin{equation}
\label{gap-eqs}
h_i +\frac{N_c}{6\pi^2} M_i\left[3I_0 - \left(3M_i^2 - M^2\right)I_1\right] = 0,
\end{equation}
with $M^2 = M_u^2 + M_d^2 + M_s^2$. Their form is unaltered by the presence of spin-1 modes in the model.

\section{Mass Diagonalization and Weak Decay Constants 
\label{s4}}

Let us consider now the free part of the Lagrangian density, which comprises the kinetic and mass terms for the boson fields. By requiring that the kinetic terms have the standard form (i.e. yielding propagators with a residue of 1 at the pole), we may determine appropriate renormalization constants for the fields. From the mass terms we are able to extract the relations between the boson masses and the model's parameters which are essential for fitting the model. The computation of axial currents is also dependent on the field renormalization constants, and may in turn be used to find expressions for the weak decay constants of the pseudoscalar mesons by applying the PCAC hypothesis. The decay constants may also be employed in the model's fitting. With this in mind, we gather the quadratic terms of the full bosonized Lagrangian density and write them out as
\begin{align}
\label{quad-Lag-1}
\mathcal{L}_{bos}^{\left(2\right)} & = \frac{1}{2} \left(h_{ab}^{\left(1\right)} \sigma_a \sigma_b + h_{ab}^{\left(2\right)} \phi_a \phi_b + H_{ab}^{\left(1\right)} V^{\mu}_a V_{b\mu} + H_{ab}^{\left(2\right)} A^{\mu}_a A_{b\mu}\right) \nonumber \\
& + \frac{N_c I_1}{16\pi^2} \tr_F{\left[\left(\partial^{\mu} \sigma\right)\left(\partial_{\mu} \sigma\right) + \left(\partial^{\mu} \phi\right)\left(\partial_{\mu} \phi\right) - \frac{1}{3} \left(F^{\mu\nu}_{\left(V\right)}F_{\mu\nu}^{\left(V\right)} + F^{\mu\nu}_{\left(A\right)}F_{\mu\nu}^{\left(A\right)}\right)\right]} \nonumber \\
& + \frac{N_c I_1}{16\pi^2} \tr_F{\left[\left[\phi,M\right]^2 - \left\lbrace \sigma, M \right\rbrace^2 - \left(\Delta_{ud} \lambda_3 + \frac{\Delta_{us} + \Delta_{ds}}{\sqrt{3}} \lambda_8\right) \left(\sigma^2 + \phi^2\right) \right.} \nonumber \\
& \left. - \left[V^{\mu},M\right] \left[V_{\mu},M\right] + \left\lbrace A^{\mu},M \right\rbrace \left\lbrace A_{\mu},M \right\rbrace \right] + \frac{N_c I_0}{8\pi^2} \tr_F{\left(\sigma^2 + \phi^2\right)} \nonumber \\
& + \frac{N_c I_1}{8\pi^2} \tr_F{\left(i \left[V^{\mu},M\right]\partial_{\mu}\sigma - \left\lbrace A^{\mu},M \right\rbrace \partial_{\mu}\phi \right)}.
\end{align}
Here, $F_{\mu\nu}^{\left(V\right)} = \partial_{\mu} V_{\nu} - \partial_{\nu} V_{\mu}$ is the field strength tensor associated with the vector field, and similarly for $F_{\mu\nu}^{\left(A\right)}$. The traces $\tr_F$ are to be taken in flavor space. From $\mathcal{L}_{SPA}$ in (\ref{SPA-Lag}) we get the four mass terms appearing in the first line of (\ref{quad-Lag-1}). Further contributions to the spin-0 mass terms stem from both the $b_1$ (the term proportional to $I_0$ in the fourth line) and $b_2$ terms in the heat kernel series in exactly the same form as in the spin-0 version of the model. From the $b_2$ term we additionally get all the kinetic terms (shown in the second line), as well as mass terms for the vectors and terms mixing spin-0 and spin-1 fields in the combinations $V^\mu\partial_\mu\sigma$ and $A^\mu\partial_\mu\phi$ (last line of (\ref{quad-Lag-1})). This mixing is a known feature arising due to spontaneous chiral symmetry breaking (for axial-vector modes) and explicit symmetry breaking (for vector modes) (see e.g. \cite{Ebert86,Osipov2002}).

In order to be able to interpret elementary excitations of the boson fields as mass eigenstates, we need an adequate redefinition of the fields which eliminate the aforementioned quadratic mixing terms. There are several possibilities here. One may use the covariant approach \cite{Osipov2002} which conserved the chiral transformation laws of spin-1 fields, or a conventional approach \cite{Osipov85,Volkov1986,Wakamatsu89,Osipov96}. The latter changes the transformation laws of spin-1 states but is simple with minimal impact on the structure of the Lagrangian and without violation of chiral symmetry. In our case, however, a shift $V_{\mu}\to V_{\mu}+kX_{\mu}$ and similarly for $A_{\mu}\to A_\mu +k'Y_\mu$, where $X_{\mu}=-i\left[M,\partial_{\mu}\sigma\right], Y_{\mu}  = \lbrace M, \partial_{\mu}\phi\rbrace,$ is not enough to achieve diagonalization. The reason is that with the inclusion of the complete set of next-to-leading order multi-quark terms, after a shift of the $V_{\mu}$ and $A_{\mu}$ fields, each combination of field components $V_a^{\mu}\partial_{\mu}\sigma_b$ (and $A_a^{\mu}\partial_{\mu}\phi_b$) will in principle need a different condition in order to be eliminated due to the complex structure of the $H_{ab}$ coefficients. This leads us to introduce shifts of the form
\begin{align}
\label{shifts}
&V_{a\mu}\rightarrow V_{a\mu}+k_a X_{a\mu}, \quad A_{a\mu}\rightarrow A_{a\mu}+k'_a Y_{a\mu}, \\
&X_{a\mu}=2f_{abc}M_b\partial_{\mu}\sigma_c, \quad Y_{a\mu}=2d_{abc}M_b\partial_{\mu}\phi_c,
\end{align}
where the index $a$ in $k_a$ and $k'_a$ is a free index. This means that each component of the fields $V^{\mu}_a$ and $A^{\mu}_a$ is independently shifted by a constant times a combination of mass and field components. Such shifts do not change the field strengths $F_{a\mu\nu}^{\left(V\right)} \to F_{a\mu\nu}^{\left(V\right)}$ and $F_{a\mu\nu}^{\left(A\right)} \to F_{a\mu\nu}^{\left(A\right)}$. 

Following the steps described in appendix \ref{app2}, we obtain for the coefficients $k_a$ related with the $V\sigma$-mixing the conditions 
\begin{align}
\label{k}
\frac{1}{k_{1,2}} & = \frac{4\pi^2 H_{11}^{\left(1\right)}}{N_c I_1} + \left(M_u - M_d\right)^2 = \frac{4\pi^2 H_{22}^{\left(1\right)}}{N_c I_1} + \left(M_u - M_d\right)^2 \nonumber \\
\frac{1}{k_{4,5}} & = \frac{4\pi^2 H_{44}^{\left(1\right)}}{N_c I_1} + \left(M_u - M_s\right)^2 = \frac{4\pi^2 H_{55}^{\left(1\right)}}{N_c I_1} + \left(M_u - M_s\right)^2 \nonumber \\
\frac{1}{k_{6,7}} & = \frac{4\pi^2 H_{66}^{\left(1\right)}}{N_c I_1} + \left(M_d - M_s\right)^2 = \frac{4\pi^2 H_{77}^{\left(1\right)}}{N_c I_1} + \left(M_d - M_s\right)^2.
\end{align}
In these expressions, we have made explicit the fact that $H_{aa}^{\left(1\right)} = H_{bb}^{\left(1\right)}$ for the pairs of indices $(a,b)=(1,2), (4,5),(6,7)$, which in turn results in the equalities $k_1 = k_2$, $k_4 = k_5$, and $k_6 = k_7$. These equalities connect pairs of components which contribute to the same $U\left(3\right)$ matrix entries and are, in light of that pattern, unsurprising. The constants $k_0,k_3,k_8$ (or alternatively $k_u,k_d,k_s$ in the flavor $\left(u,d,s\right)$ basis) remain unconstrained by the diagonalization requirements, since mixing terms of field components associated with them are already zero due to the well-known properties of the antisymmetric constants $f_{abc}$. Therefore, we may choose $k_a = 0$, $a=0,3,8$.

To avoid the $A\phi$-mixing we come to the following conditions:
\begin{align*}
\label{k'}
\frac{1}{k'_{1,2}} & = \frac{4\pi^2 H_{11}^{\left(2\right)}}{N_c I_1} + \left(M_u + M_d\right)^2 = \frac{4\pi^2 H_{22}^{\left(2\right)}}{N_c I_1} + \left(M_u + M_d\right)^2 \nonumber \\
\frac{1}{k'_{4,5}} & = \frac{4\pi^2 H_{44}^{\left(2\right)}}{N_c I_1} + \left(M_u + M_s\right)^2 = \frac{4\pi^2 H_{55}^{\left(2\right)}}{N_c I_1} + \left(M_u + M_s\right)^2 \nonumber \\
\end{align*}
\begin{align}
&\frac{1}{k'_{6,7}} = \frac{4\pi^2 H_{66}^{\left(2\right)}}{N_c I_1} + \left(M_d + M_s\right)^2 = \frac{4\pi^2 H_{77}^{\left(2\right)}}{N_c I_1} + \left(M_d + M_s\right)^2 \nonumber \\
&\frac{1}{k'_{u,d,s}} = \frac{8\pi^2 H_{uu,dd,ss}^{\left(2\right)}}{N_c I_1} + 4M_{u,d,s}^2.
\end{align}
As in the previous case, we recognize equalities $k'_1 = k'_2$, $k'_4 = k'_5$, and $k'_6 = k'_7$, which rely on equivalent equalities among $H_{aa}^{\left(2\right)}$ components. Furthermore, we have defined the constants $k'_i$ in the flavor basis $i=u,d,s$ as $k'_i = 2 \sum_a k'_a e_{ai}^2$, and $H_{ij} = \sum_{a,b} H_{ab} e_{ai} e_{bj}$. It should be noted that $2H_{uu,dd} = H_{11,22}$ in the isospin limit, whereas $2H_{ii} = H_{aa}$ for all $i \in \left\lbrace u,d,s \right\rbrace$, $a \neq 0,3,8$ in the full degenerate case ($m_u = m_d = m_s$).

Besides dealing with the mixing terms in the quadratic part of the bosonized Lagrangian, the shifts (\ref{shifts}) contribute to the kinetic terms of the spin-0 fields which are expressed in (\ref{quad-Lag-2}). With the use of conditions (\ref{diag-conds}), as well as (\ref{k}) and (\ref{k'}), the full kinetic terms of these fields may be simplified to
\begin{align}
\mathcal{L}_{\sigma}^{\left(kin\right)} & = \frac{N_c I_1}{8\pi^2} \partial^{\mu} \sigma_a \partial_{\mu} \sigma_b \left(\delta_{ab} + 4k_e f_{ace} f_{bde} M_c M_d\right) \nonumber \\
& = \frac{N_c I_1}{16\pi^2} \sum_{i=u,d,s} \partial^{\mu} \sigma_i \partial_{\mu} \sigma_i + \sum_{a \neq 0,3,8} \frac{k_a}{2} H_{aa}^{\left(1\right)} \partial^{\mu} \sigma_a \partial_{\mu} \sigma_a \nonumber \\
\mathcal{L}_{\phi}^{\left(kin\right)} & = \frac{N_c I_1}{8\pi^2} \partial^{\mu} \phi_a \partial_{\mu} \phi_b \left(\delta_{ab} - 4k'_e d_{ace} d_{bde} M_c M_d\right) \nonumber \\
& = \sum_{i=u,d,s} \frac{k'_i}{2} H_{ii}^{\left(2\right)} \partial^{\mu} \phi_i \partial_{\mu} \phi_i + \sum_{a \neq 0,3,8} \frac{k'_a}{2} H_{aa}^{\left(2\right)} \partial^{\mu} \phi_a \partial_{\mu} \phi_a.
\end{align}
These expressions lead us to define the necessary field rescalings for obtaining kinetic terms in their standard forms. These are
\begin{align}
\label{sigma-phi-rescale}
\sigma_i & \rightarrow\sqrt{\frac{4\pi^2}{N_c I_1}}\sigma_i = \varrho \sigma_i \quad \left(i=u,d,s\right),\nonumber \\
\sigma_a & \rightarrow \sqrt{\frac{1}{k_a H_{aa}^{\left(1\right)}}} \sigma_a = \varrho \sqrt{1 + \frac{\xi_a^{\sigma}}{\varrho^2}} \sigma_a \quad \left(a \neq 0,3,8\right), \nonumber \\
\phi_i & \rightarrow \sqrt{\frac{1}{2k'_i H_{ii}^{\left(2\right)}}} \phi_i =\varrho \sqrt{1 + \frac{\xi_i^{\phi}}{\varrho^2}} \phi_i \quad \left(i=u,d,s\right), \nonumber \\
\phi_a & \rightarrow \sqrt{\frac{1}{k'_a H_{aa}^{\left(1\right)}}} \phi_a = \varrho \sqrt{1 + \frac{\xi_a^{\phi}}{\varrho^2}} \phi_a \quad \left(a \neq 0,3,8\right),
\end{align}
\noindent with $\varrho^2 = 4\pi^2/(N_c I_1)$ and
\begin{eqnarray}
\xi_1^{\sigma}&=&\xi_2^{\sigma}=\frac{\left(M_u - M_d\right)^2}{H_{11}^{\left(1\right)}}, \quad \xi_1^{\phi}=\xi_2^{\phi}=\frac{\left(M_u + M_d\right)^2}{H_{11}^{\left(2\right)}}, \nonumber \\
\xi_4^{\sigma}&=&\xi_5^{\sigma}=\frac{\left(M_u - M_s\right)^2}{H_{44}^{\left(1\right)}},\quad \xi_4^{\phi}=\xi_5^{\phi}=\frac{\left(M_u + M_s\right)^2}{H_{44}^{\left(2\right)}}, \nonumber \\
\xi_6^{\sigma}&=&\xi_7^{\sigma}=\frac{\left(M_d - M_s\right)^2}{H_{66}^{\left(1\right)}}, \quad\xi_6^{\phi}=\xi_7^{\phi}=\frac{\left(M_d + M_s\right)^2}{H_{66}^{\left(2\right)}}, \nonumber \\
&&\ \ \ \ \ \ \ \ \ \ \ \ \ \ \ \ \ \ \ \ \ \ \ \ \ \ \ \, \xi_i^{\phi}=\frac{2 M_i^2}{H_{ii}^{\left(2\right)}}.
\end{eqnarray}
\noindent One should keep in mind the relations between $H_{ii}$ and $H_{aa}$ as stated after equation (\ref{k'}).

In exactly the same way, one may extract the kinetic terms of spin-1 fields in (\ref{quad-Lag-2}),
\begin{equation}
\mathcal{L}_V^{\left(kin\right)}= -\frac{N_c I_1}{48\pi^2} \tr_F{\left(F^{\mu\nu}_{\left(V\right)} F_{\mu\nu}^{\left(V\right)}\right)}, \quad
\mathcal{L}_A^{\left(kin\right)} = -\frac{N_c I_1}{48\pi^2} \tr_F{\left(F^{\mu\nu}_{\left(A\right)} F_{\mu\nu}^{\left(A\right)}\right)},
\end{equation}
and rescale them as
\begin{align}
\label{V-A-rescale}
V^{\mu}_a \rightarrow \sqrt{\frac{3}{2}} \varrho V^{\mu}_a, \nonumber \\
A^{\mu}_a \rightarrow \sqrt{\frac{3}{2}} \varrho A^{\mu}_a,
\end{align}
in order to obtain standard kinetic terms for spin-1 fields.

Finally, we apply the rescalings defined in (\ref{sigma-phi-rescale}) and (\ref{V-A-rescale}) to the remaining terms in (\ref{quad-Lag-2}) so that we may rewrite the quadratic part of the bosonized Lagrangian density as
\begin{equation}
\label{total-quad-Lag}
\mathcal{L}_{bos}^{\left(2\right)} = \mathcal{L}_{\sigma}^{\left(2\right)} + \mathcal{L}_{\phi}^{\left(2\right)} + \mathcal{L}_{V}^{\left(2\right)} + \mathcal{L}_{A}^{\left(2\right)},
\end{equation}
with
\begin{align}
\label{quad-Lag-sigma-2}
\mathcal{L}_{\sigma}^{\left(2\right)} & = \frac{1}{4} \sum_{i=u,d,s} \partial^{\mu} \sigma_i \partial_{\mu} \sigma_i + \frac{1}{2} \sum_{a \neq 0,3,8} \partial^{\mu} \sigma_a \partial_{\mu} \sigma_a \nonumber \\
& + \frac{1}{4} \sum_{i=u,d,s} \sigma_i^2 \left[ \varrho^2 \left(2 h_{ii}^{\left(1\right)} - \frac{h_i}{M_i} \right) - 4M_i^2 \right] + \frac{1}{2} \sum_{i \neq j} \sigma_i \sigma_j \varrho^2 h_{ij}^{\left(1\right)} \nonumber \\
& + \frac{1}{2} \left(\sigma_1^2 + \sigma_2^2\right) \left(\varrho^2 + \frac{\left(M_u - M_d\right)^2}{H_{11}^{\left(1\right)}} \right) \left(h_{11}^{\left(1\right)} - \frac{h_u - h_d}{M_u - M_d}\right) \nonumber \\
& + \frac{1}{2} \left(\sigma_4^2 + \sigma_5^2\right) \left(\varrho^2 + \frac{\left(M_u - M_s\right)^2}{H_{44}^{\left(1\right)}} \right) \left(h_{44}^{\left(1\right)} - \frac{h_u - h_s}{M_u - M_s}\right) \nonumber \\
& + \frac{1}{2} \left(\sigma_6^2 + \sigma_7^2\right) \left(\varrho^2 + \frac{\left(M_d - M_s\right)^2}{H_{66}^{\left(1\right)}} \right) \left(h_{66}^{\left(1\right)} - \frac{h_d - h_s}{M_d - M_s}\right),
\end{align}
\begin{align}
\label{quad-Lag-phi-2}
\mathcal{L}_{\phi}^{\left(2\right)} & = \frac{1}{4} \sum_{i=u,d,s} \partial^{\mu} \phi_i \partial_{\mu} \phi_i + \frac{1}{2} \sum_{a \neq 0,3,8} \partial^{\mu} \phi_a \partial_{\mu} \phi_a \nonumber \\
& + \frac{1}{4} \sum_{i=u,d,s} \phi_i^2 \left(\varrho^2 + \frac{2 M_i^2}{H_{ii}^{\left(2\right)}} \right) \left( 2h_{ii}^{\left(2\right)} - \frac{h_i}{M_i} \right) \nonumber \\
& + \frac{1}{2} \sum_{i \neq j} \phi_i \phi_j \sqrt{\left(\varrho^2 + \frac{2 M_i^2}{H_{ii}^{\left(2\right)}} \right) \left(\varrho^2 + \frac{2 M_i^2}{H_{jj}^{\left(2\right)}}\right)} h_{ij}^{\left(2\right)} \nonumber \\
& + \frac{1}{2} \left(\phi_1^2 + \phi_2^2\right) \left(\varrho^2 + \frac{\left(M_u + M_d\right)^2}{H_{11}^{\left(2\right)}} \right) \left( h_{11}^{\left(2\right)} - \frac{h_u + h_d}{M_u + M_d} \right) \nonumber \\
& + \frac{1}{2} \left(\phi_4^2 + \phi_5^2\right) \left(\varrho^2 + \frac{\left(M_u + M_s\right)^2}{H_{44}^{\left(2\right)}} \right) \left( h_{44}^{\left(2\right)} - \frac{h_u + h_s}{M_u + M_s} \right) \nonumber \\
& + \frac{1}{2} \left(\phi_6^2 + \phi_7^2\right) \left(\varrho^2 + \frac{\left(M_d + M_s\right)^2}{H_{66}^{\left(2\right)}} \right) \left( h_{66}^{\left(2\right)} - \frac{h_d + h_s}{M_d + M_s} \right),
\end{align}
\begin{align}
\label{quad-Lag-V}
\mathcal{L}_{V}^{\left(2\right)} & = -\frac{1}{8} \sum_{i=u,d,s} F^{\mu\nu}_{i\left(V\right)} F_{i\mu\nu}^{\left(V\right)} - \frac{1}{4} \sum_{a \neq 0,3,8} F^{\mu\nu}_{a\left(V\right)} F_{a\mu\nu}^{\left(V\right)} + \frac{3}{4} \sum_{i=u,d,s} V^{\mu}_i V_{i\mu} \varrho^2 H_{ii}^{\left(1\right)} \nonumber \\
& + \frac{3}{4} \left(V^{\mu}_1 V_{1\mu} + V^{\mu}_2 V_{2\mu}\right) \varrho^2 \left[H_{11}^{\left(1\right)} + \frac{N_c I_1}{4\pi^2} \left(M_u - M_d\right)^2\right] \nonumber \\
& + \frac{3}{4} \left(V^{\mu}_4 V_{4\mu} + V^{\mu}_5 V_{5\mu}\right) \varrho^2 \left[H_{44}^{\left(1\right)} + \frac{N_c I_1}{4\pi^2} \left(M_u - M_s\right)^2\right] \nonumber \\
& + \frac{3}{4} \left(V^{\mu}_6 V_{6\mu} + V^{\mu}_7 V_{7\mu}\right) \varrho^2 \left[H_{66}^{\left(1\right)} + \frac{N_c I_1}{4\pi^2} \left(M_d - M_s\right)^2\right]
\end{align}
and
\begin{align}
\label{quad-Lag-A}
\mathcal{L}_{A}^{\left(2\right)} & = -\frac{1}{8} \sum_{i=u,d,s} F^{\mu\nu}_{i\left(A\right)} F_{i\mu\nu}^{\left(A\right)} - \frac{1}{4} \sum_{a \neq 0,3,8} F^{\mu\nu}_{a\left(A\right)} F_{a\mu\nu}^{\left(A\right)} \nonumber \\
& + \frac{3}{4} \sum_{i=u,d,s} A^{\mu}_i A_{i\mu} \varrho^2 \left( H_{ii}^{\left(2\right)} + \frac{N_c I_1}{2\pi^2} M_i^2 \right) \nonumber \\
& + \frac{3}{4} \left(A^{\mu}_1 A_{1\mu} + A^{\mu}_2 A_{2\mu}\right) \varrho^2 \left[H_{11}^{\left(2\right)} + \frac{N_c I_1}{4\pi^2} \left(M_u + M_d\right)^2\right] \nonumber \\
& + \frac{3}{4} \left(A^{\mu}_4 A_{4\mu} + A^{\mu}_5 A_{5\mu}\right) \varrho^2 \left[H_{44}^{\left(2\right)} + \frac{N_c I_1}{4\pi^2} \left(M_u + M_s\right)^2\right] \nonumber \\
& + \frac{3}{4} \left(A^{\mu}_6 A_{6\mu} + A^{\mu}_7 A_{7\mu}\right) \varrho^2 \left[H_{66}^{\left(2\right)} + \frac{N_c I_1}{4\pi^2} \left(M_d + M_s\right)^2\right].
\end{align}
We have used the gap equations (\ref{gap-eqs}) to simplify the scalar (\ref{quad-Lag-sigma-2}) and pseudoscalar (\ref{quad-Lag-phi-2}) field quadratic terms. The terms proportional to $h_{ij}$ in both (\ref{quad-Lag-sigma-2}) and (\ref{quad-Lag-phi-2}) express the well-known mixings occurring between the flavor components forming the neutral scalar and pseudoscalar mesons. Moreover, these expressions reveal the impact of the new parameters $w_i$ in the spin-0 sector masses through the $H_{aa}$ coefficients. These modifications are a direct consequence of $V\sigma$- and $A\phi$-mixing in the bosonized Lagrangian in (\ref{quad-Lag-1}), and they enter all mass terms except those of the neutral scalar sector. 

We collect the nine $(a=0,1,\ldots,8)$ scalar $(\sigma_a)$, pseudoscalar $(\phi_a)$, vector $(V^\mu_a)$ and axial-vector $(A^\mu_a)$ fields in the hermitian matrices 
\begin{align}
\frac{\sigma}{\sqrt{2}} & = \left( 
\begin{array}{ccc}
\frac{\sigma_u}{\sqrt{2}} & a_0^{+} & \kappa^{+} \\
a_0^{-} & \frac{\sigma_d}{\sqrt{2}} & \kappa^0 \\
\kappa^{-} & \bar{\kappa}^0 & \frac{\sigma_s}{\sqrt{2}}
\end{array} \right), 
\qquad \quad \quad
\frac{\phi}{\sqrt{2}}= \left( \begin{array}{ccc}
\frac{\phi_u}{\sqrt{2}} & \pi^{+} & K^{+} \\
\pi^{-} & \frac{\phi_d}{\sqrt{2}} & K^0 \\
K^{-} & \bar{K}^0 & \frac{\phi_s}{\sqrt{2}}
\end{array} \right), \\
\frac{V_{\mu}}{\sqrt{2}} & = \left( \begin{array}{ccc}
\frac{V_{u \mu}}{\sqrt{2}} & \rho_{\mu}^{+} & K_{\mu}^{*+} \\
\rho_{\mu}^{-} & \frac{V_{d \mu}}{\sqrt{2}} & K_{\mu}^{*0} \\
K_{\mu}^{*-} & \bar{K}_{\mu}^{*0} & \frac{V_{s \mu}}{\sqrt{2}}
\end{array} \right), \qquad
\frac{A_{\mu}}{\sqrt{2}} = \left( \begin{array}{ccc}
\frac{A_{u \mu}}{\sqrt{2}} & a_{1\mu}^{+} & K_{1\mu}^{+} \\
a_{1\mu}^{-} & \frac{A_{d \mu}}{\sqrt{2}} & K_{1\mu}^{0} \\
K_{1\mu}^{-} & \bar{K}_{1\mu}^{0} & \frac{A_{s \mu}}{\sqrt{2}}
\end{array} \right).
\end{align}
Here the flavor basis boson fields may be represented as a linear combinations of $a=0,3,8$ states 
\begin{equation}
\label{neutral-sigma-sector}
\!\!\!\!\!\!\!\!\!\!\!\!\!\!\!\!\!\!\!\!\!\!\!\!\!
\left\lbrace 
\begin{array}{lcl} \sigma_u & = & \sigma_3 + \frac{\sqrt{2}\sigma_0 + \sigma_8}{\sqrt{3}} = \sigma_3 + f_{ns} \\
\sigma_d & = & - \sigma_3 + \frac{\sqrt{2}\sigma_0 + \sigma_8}{\sqrt{3}} = - \sigma_3 + f_{ns} \\
\sigma_s & = & \frac{\sqrt{2}\sigma_0-2\sigma_8}{\sqrt{3}}=\sqrt{2} f_s 
\end{array} \right.
\end{equation}
\begin{equation}
\label{neutral-phi-sector}
\!\!\!\!\!\!\!\!\!\!\!\!\!\!\!\!\!\!\!\!\!\!\!\!\!
\left\lbrace \begin{array}{lcl} \phi_u & = & \phi_3 + \frac{\sqrt{2}\phi_0 + \phi_8}{\sqrt{3}} = \phi_3 + \eta_{ns} \\
\phi_d & = & - \phi_3 + \frac{\sqrt{2}\phi_0 + \phi_8}{\sqrt{3}} = - \phi_3 + \eta_{ns} \\
\phi_s & = & \frac{\sqrt{2}\phi_0 -2 \phi_8}{\sqrt{3}} = \sqrt{2} \eta_s \end{array} \right.
\end{equation}
\begin{equation}
\label{neutral-V-sector}
\!\!\!\!\!\!\!\!\!\!\!\!\!\!\!\!\!
\left\lbrace \begin{array}{lcl} V^\mu_{u} & = & V^\mu_{3} + \frac{\sqrt{2}V^\mu_{0} + V^\mu_{8}}{\sqrt{3}} = V^\mu_{3} + \omega^\mu_{ns} \\
V^\mu_{d} & = & - V^\mu_{3} + \frac{\sqrt{2}V^\mu_{0} + V^\mu_{8}}{\sqrt{3}} = - V^\mu_{3} + \omega^\mu_{ns} \\
V^\mu_{s} & = & \frac{\sqrt{2}V^\mu_{0} -2 V^\mu_{8}}{\sqrt{3}}= \sqrt{2} \omega^\mu_{s} \end{array} \right.
\end{equation}
\begin{equation}
\label{neutral-A-sector}
\!\!\!\!\!\!\!\!\!\!\!\!
\left\lbrace \begin{array}{lcl} A^\mu_{u} & = & A^\mu_{3} + \frac{\sqrt{2}A^\mu_{0} + A^\mu_{8}}{\sqrt{3}} = A^\mu_{3} + f^\mu_{1ns} \\
A^\mu_{d} &=& - A^\mu_{3} + \frac{\sqrt{2}A^\mu_{0} + A^\mu_{8}}{\sqrt{3}} = - A^\mu_{3} + f^\mu_{1ns} \\
A^\mu_{s} & = & \frac{\sqrt{2}A^\mu_{0} -2 A^\mu_{8}}{\sqrt{3}} = \sqrt{2} f^\mu_{1s} \end{array} \right. .
\end{equation}
In relations (\ref{neutral-sigma-sector})-(\ref{neutral-A-sector}), we have decomposed the diagonal components of the fields in the $\left(0,3,8\right)$ basis, as well as in the $\left(3,ns,s\right)$ basis, which consists of the neutral isotriplet component and the non-strange and strange isosinglet components, respectively. These three components still appear in bilinear mixed terms in (\ref{quad-Lag-sigma-2}) and (\ref{quad-Lag-phi-2}) due to the term proportional to $h_{ij}$. 

The physical (mass eigenstates) neutral mesons arise as suitable combinations of said components which may be parametrized by three mixing angles constrained by three diagonalization conditions imposed by the requirement that all mixing terms are eliminated. In the isospin limit, the neutral isotriplet component uncouples from the isosinglet ones, and only one mixing angle is needed for diagonalization. We may then identify immediately $\sigma_3=a_0^0$, $\phi_3=\pi^0$, $V_{3\mu}=\rho_\mu^0$ and $A_{3\mu}=a_{1\mu}^0$. In the pseudoscalar sector the mixing angle is commonly introduced as
\begin{equation}
\label{theta-phi-angle}
\left( \begin{array}{c} \eta \\ \eta' \end{array} \right) = \left( \begin{array}{cc} \cos{\theta_{\phi}} & -\sin{\theta_{\phi}} \\ \sin{\theta_{\phi}} & \cos{\theta_{\phi}} \end{array} \right) \left( \begin{array}{c} \phi_8 \\ \phi_0 \end{array} \right)
\end{equation}
in the $\left(0,8\right)$ basis or as
\begin{equation}
\label{psi-phi-angle}
\left( \begin{array}{c} \eta \\ \eta' \end{array} \right) = \left( \begin{array}{cc} \cos{\psi_{\phi}} & -\sin{\psi_{\phi}} \\
\sin{\psi_{\phi}} & \cos{\psi_{\phi}}
\end{array} \right) \left( \begin{array}{c} \eta_{ns} \\ \eta_s \end{array} \right)
\end{equation}
in the $\left(ns,s\right)$ basis. In the scalar sector, the mixing scheme is analogous to that of the pseudoscalar sector,
\begin{equation}
\label{theta-sigma-angle}
\left( 
\begin{array}{c} f_0 \\ \sigma 
\end{array} \right) = \left( 
\begin{array}{cc} \cos{\theta_{\sigma}} & -\sin{\theta_{\sigma}} \\ \sin{\theta_{\sigma}} & \cos{\theta_{\sigma}} 
\end{array} \right) \left( 
\begin{array}{c} \sigma_8 \\ \sigma_0 
\end{array} \right)
\end{equation}
or
\begin{equation}
\label{psi-sigma-angle}
\left( \begin{array}{c} f_0 \\ \sigma \end{array} \right) = \left( \begin{array}{cc} \cos{\psi_{\sigma}} & -\sin{\psi_{\sigma}} \\
\sin{\psi_{\sigma}} & \cos{\psi_{\sigma}}
\end{array} \right) \left( \begin{array}{c} f_{ns} \\ f_s \end{array} \right).
\end{equation}
In both cases, the $\theta$ and $\psi$ angles are related to each other through $\psi = \theta + \arctan{\sqrt{2}}$. The resulting scalar and pseudoscalar mass Lagrangians are listed in Appendix \ref{app3}, together with the vector and axial vector ones (which involve no mixing between the neutral mesons). We may then summarize the mass expressions. For scalar mesons we have
\begin{align*}
\label{scalar-masses}
M_{a_0}^2 & = \varrho^2 \left(\frac{h_{u}}{M_{u}}- h_{11}^{\left(1\right)}\right) + 4 M_{u}^2, \nonumber \\
M_{\kappa}^2 & = \left(\varrho^2 + \frac{\left(M_{u} - M_s \right)^2}{H_{44}^{\left(1\right)}} \right) \left(\frac{h_{u} - h_s}{M_{u} - M_s} - h_{44}^{\left(1\right)}\right), \nonumber \\
\end{align*}
\begin{align}
M_{f_0}^2&=\frac{1}{1-\tan^2{\psi_{\sigma}}}\left[\varrho^2\left(\frac{h_{u}}{M_{u}}-2h_{uu}^{\left(1\right)}-2h_{ud}^{\left(1\right)}\right)+4M_{u}^2\right]+\frac{1}{1-\cot^2{\psi_{\sigma}}}\left[\varrho^2\left(\frac{h_s}{M_s}-2h_{ss}^{\left(1\right)}\right)+4M_s^2\right], \nonumber \\
M_{\sigma}^2&=\frac{1}{1-\cot^2{\psi_{\sigma}}}\left[\varrho^2\left(\frac{h_{u}}{M_{u}}-2h_{uu}^{\left(1\right)}-2h_{ud}^{\left(1\right)}\right)+4M_{u}^2\right]+\frac{1}{1-\tan^2{\psi_{\sigma}}}\left[\varrho^2\left(\frac{h_s}{M_s}-2h_{ss}^{\left(1\right)}\right)+4M_s^2\right].
\end{align}
The masses of the pseudoscalars are given by
\begin{align}
\label{pseudoscalar-masses}
M_{\pi}^2 & = \left(\varrho^2 + \frac{4 M_{u}^2}{H_{11}^{\left(2\right)}}\right) \left(\frac{h_{u}}{M_{u}}- h_{11}^{\left(2\right)}\right),\nonumber \\
M_K^2 & = \left(\varrho^2 + \frac{\left(M_{u} + M_s \right)^2}{H_{44}^{\left(2\right)}} \right) \left(\frac{h_{u} + h_s}{M_{u} + M_s} - h_{44}^{\left(2\right)}\right),\nonumber \\
M_{\eta}^2 & = \frac{1}{1-\tan^2{\psi_{\phi}}} \left(\varrho^2 + \frac{2 M_{u}^2}{H_{uu}^{\left(2\right)}}\right) \left(\frac{h_{u}}{M_{u}} - 2 h_{uu}^{\left(2\right)} - 2 h_{ud}^{\left(2\right)}\right) + \frac{1}{1-\cot^2{\psi_{\phi}}} \left(\varrho^2 + \frac{2 M_s^2}{H_{ss}^{\left(2\right)}}\right) \left(\frac{h_s}{M_s} - 2 h_{ss}^{\left(2\right)}\right), \nonumber \\
M_{\eta'}^2 & = \frac{1}{1-\cot^2{\psi_{\phi}}} \left(\varrho^2 + \frac{2 M_{u}^2}{H_{uu}^{\left(2\right)}}\right) \left(\frac{h_{u}}{M_{u}} - 2 h_{uu}^{\left(2\right)} - 2 h_{ud}^{\left(2\right)}\right) + \frac{1}{1-\tan^2{\psi_{\phi}}} \left(\varrho^2 + \frac{2 M_s^2}{H_{ss}^{\left(2\right)}}\right) \left(\frac{h_s}{M_s} - 2 h_{ss}^{\left(2\right)}\right).
\end{align}
In the chiral limit ($m_i \to 0$), we get $M_u = M_s = M$, $h_u = h_s = h$ and $h_{aa}^{\left(2\right)} = h^{\left(2\right)}$, as well as $h = Mh^{\left(2\right)}$. This makes it easy to verify in (\ref{pseudoscalar-masses}) that both $M_{\pi}$ and $M_K$ go to zero in this limit. Also, in the chiral limit we get $h^{\left(2\right)}_{ii} = h^{\left(2\right)'}_1$ and $h^{\left(2\right)}_{ij} = h^{\left(2\right)'}_2$ ($i\neq j$), with $2\left(h^{\left(2\right)'}_1 - h^{\left(2\right)'}_2\right) = h^{\left(2\right)}$. Using this and the fact that $\psi_{\phi} = \arctan{\sqrt{2}}$ in this limit (this may be checked resorting to the conditions \ref{angle-conds} and is equivalent to $\theta_{\phi} = 0$, i.e. no mixing between flavor $SU\left(3\right)$ singlet $0$ and octet $8$ components), it can be shown that $M_{\eta}$ also goes to zero, while the $\eta'$ retains a finite mass, due to the Adler-Bell-Jackiw anomaly, given by
\begin{equation}
M_{\eta'}^2 \to \left(\varrho^2 + \frac{2M^2}{H_{ii}^{\left(2\right)}}\right) \frac{2}{2G + \left(3g_1 h + g_2 h + \kappa\right)h}.
\end{equation}
\noindent A detailed discussion of the anomaly within the $8q$-extended version of the model without the explicit symmetry breaking interactions and the vector terms is given in \cite{Osipov2007}.

For the masses of the vector and axial-vector mesons we obtain
\begin{align}
\label{vector-masses}
&M_{\rho}^2=M_{\omega}^2=\frac{3}{2}\varrho^2 H_{11}^{\left(1\right)}, \nonumber \\
&M_{K^*}^2=\frac{3}{2}\left[\varrho^2 H_{44}^{\left(1\right)}+\left(M_{u}-M_s\right)^2 \right], \nonumber \\
&M_{\varphi}^2=3\varrho^2H_{ss}^{\left(1\right)}, 
\end{align}

\begin{align}
\label{axial-masses}
&M_{a_1}^2= M_{f_1}^2=\frac{3}{2}\varrho^2H_{11}^{\left(2\right)}+6 M_{u}^2, \nonumber \\
&M_{K_1}^2=\frac{3}{2}\left[\varrho^2 H_{44}^{\left(2\right)}+\left(M_{u}+M_s\right)^2\right], \nonumber \\
&M_{f'_1}^2= 3 \varrho^2H_{ss}^{\left(2\right)}+6 M_s^2.
\end{align}

Axial transformations of the meson fields may be used \cite{Levi60} to define axial vector currents $\mathcal{A}_a^{\mu}$, which are conserved in the chiral limit ($m_i = 0,\, i=u,d,s$). The linear (in powers of meson fields) part of these currents may be obtained from (\ref{total-quad-Lag}). The general formula is
\begin{equation}
\mathcal{A}_a^{\mu} = \frac{\partial \delta \mathcal{L}}{\partial\left(\partial_{\mu}\beta_a\right)} = \sum_C \frac{\partial \mathcal{L}}{\partial\left(\partial_{\nu} C\right)} \frac{\partial \left(\partial_{\nu}\delta C\right)}{\partial\left(\partial_{\mu} \beta_a\right)},
\end{equation}
where $\delta\mathcal{L}$ is the variation of the Lagrangian due to the local axial transformations parametrized by $\beta_a(x)$. The sum in $C$ is over all $\sigma$, $\phi$, $V_\mu$ and $A_\mu$ field components, and $\delta C$ is the infinitesimal chiral transformation of the field $C$. Having this current one may wish to calculate the matrix element $\left\langle 0\right|\mathcal{A}_{a\mu}\left|\phi_b\left(p\right)\right\rangle = -if_{ab}p_{\mu}$, where the $f_{ab}$ are constants associated with the weak decays of pseudoscalar states. For the pion weak decay constant $f_{\pi}$ we find
\begin{equation}
\left\langle 0 \right| \frac{1}{\sqrt{2}}\left(\mathcal{A}_{1\mu}\pm i\mathcal{A}_{2\mu}\right)\left|\pi^{\mp}\left(p\right)\right\rangle =-if_{\pi}p_{\mu},
\end{equation}
where $f_\pi$ (in the isospin limit considered) is given by
\begin{equation}
f_{\pi} = \frac{M_u}{\sqrt{\varrho^2+\frac{4 M_u^2}{H_{11}^{\left(2\right)}}}}\equiv\frac{M_u}{g_\pi}.
\end{equation}
This is nothing else than the quark analog of the celebrated Goldberger-Treiman relation, which is $M_u=f_\pi g_\pi$, where $g_\pi$ is the renormalization constant of the pion field. 

In full analogy with calculations of $f_{\pi}$, the weak decay constant of the kaons can be shown to have the form
\begin{equation}
f_{K}=\frac{M_u+M_s}{2\sqrt{\varrho^2+\frac{\left(M_u+M_s\right)^2}{H_{44}^{\left(2\right)}}}}\equiv\frac{M_u+M_s}{2g_K}.
\end{equation}This expression again follows the general framework of the Goldberger-Treiman result for this quantity, $(M_u + M_s)=2f_Kg_K$, where $g_K$ is the renormalization constant of the kaon field defined through (\ref{sigma-phi-rescale}).

\section{Parameter Fitting and Discussion 
\label{s5}}

The problem of fitting the model's parameters has been thoroughly discussed in \cite{model1,model2} in the isospin limit, where only spin-0 modes had been considered. There, the number of conditions and empirical inputs was just right for the fitting to be accomplished. The inclusion of spin-1 modes adds 13 new parameters to the model while also altering the formulae for the spin-0 meson masses and weak decay constants. From the 13 new parameters, only 9 appear in the quadratic part of the bosonized Lagrangian through $H_{ab}^{\left(1,2\right)}$, as can be seen in expressions (\ref{Hab1}) and (\ref{Hab2}). (The other parameters may contribute at finite densities, at which $\left\langle V_i^{0} \right\rangle \neq 0$.) Even so, in the isospin limit, the only new available empirical inputs are the 6 independent vector and axial-vector meson masses, which are still not enough for an unambiguous fitting. If we cannot definitely pinpoint all the new parameters using only the quadratic part of the Lagrangian, we then choose to address a slightly different problem: can we find a parameter set which reproduces the full spin-0 and spin-1 meson spectra?

In total, there are 29 adjustable parameters: 2 current quark masses ($m_{u,d}$, $m_s$); 2 quark condensates ($h_{u,d}$, $h_s$); 2 constituent quark masses ($M_{u,d}$, $M_s$); 1 cutoff ($\Lambda$); 11 "old" couplings ($G$, $\kappa$, $\kappa_2$, $g_1$, $g_2$, $g_3$, $g_4$, $g_5$, $g_6$, $g_7$, $g_8$); 9 "new" couplings ($w_1$, $w_6$, $w_7$, $w_8$, $w_9$, $w_{10}$, $w_{11}$, $w_{12}$, $w_{13}$); and 2 mixing angles ($\theta_{\sigma}$, $\theta_{\phi}$). On the other hand, there are a number of conditions and empirical inputs which can be used to fit the model: 2 gap equations (\ref{gap-eqs}); 2 stationary phase conditions (\ref{hi-conds}); 4 pseudoscalar masses $(M_\pi, M_K, M_\eta, M_{\eta'})$; 4 scalar masses $(M_\sigma. M_\kappa, M_{a_0}, M_{f_0})$; 3 vector masses $(M_\rho, M_{K^*}, M_\varphi)$; 3 axial-vector masses $(M_{a_1}, M_{K_1}, M_{f_1})$; 2 pseudoscalar weak decay constants $(f_\pi, f_K)$; and 2 mixing angle conditions. These give a total of 22 conditions, which is 7 conditions short for a complete unequivocal fitting of all the parameters. 

Three of the missing conditions may be provided by externally fixing the current quark masses $m_u$ and $m_s$, as well as the pseudoscalar mixing angle $\theta_{\phi}$, to be in accord with known phenomenological expectations. 

We may also note that all $w_i$ parameters enter all expressions related with the quadratic part of the Lagrangian through $H_{ab}$ coefficients only. In (\ref{Hab1}) and (\ref{Hab2}), we see that $w_1$, $w_6$ and $w_9$ contribute in exactly the same way to all coefficients independently of $a$ and $b$, so that we may effectively set two of these parameters to zero and take only one of them to contribute for the three; we then choose $w_6 = w_9 = 0$. The correct distribution of this contribution among the three parameters may require non-zero $w_6$ and $w_9$, but this only becomes relevant when looking at interaction terms, where the $H_{abc}$ coefficients appear, and does not invalidate the idea of setting them to zero in order to attempt fitting at the quadratic Lagrangian level only.

From among the other parameters, we choose $w_{13} = 0$, but we must show that this choice is essentially arbitrary. To that end, we have used this prescription for fitting the model, and then have independently varied the fixed values for $w_1$ and $w_{13}$ and repeated the fitting. We have found that varying either $w_1$ or $w_{13}$ simply resulted in a refitting of the values of $w_8$ and $w_{11}$, with no impact on any other parameter. 

The parameters $w_7$, $w_{10}$ and $w_{12}$, which appear with opposite signals in (\ref{Hab1}) and (\ref{Hab2}), are essential for establishing the mass differences between flavor partner vector and axial-vector mesons and should therefore not be set to zero. Regarding this statement, we may actually prove the following conditions which are valid in the case of an exact isospin symmetry:
\begin{align}
&w_7 h_u^2 + 2w_{10} m_u h_u + w_{12} m_u^2 = 3 \varrho^2 \left(\frac{1}{M_{a_1}^2 - 6 M_u^2} - \frac{1}{M_{\rho}^2}\right), \nonumber \\
&w_7 h_s^2 + 2w_{10} m_s h_s + w_{12} m_s^2 = 3 \varrho^2 \left(\frac{1}{M_{f_1}^2 - 6 M_s^2} - \frac{1}{M_{\varphi}^2}\right), \nonumber \\
&w_7 h_u h_s + w_{10} \left(m_u h_s + m_s h_u\right) + w_{12} m_u m_s \nonumber \\ 
& = 3 \varrho^2 \left[\frac{1}{M_{K_1}^2 - \frac{3}{2} \left(M_u + M_s\right)^2} - \frac{1}{M_{K^*}^2 - \frac{3}{2} \left(M_u - M_s\right)^2}\right].
\end{align}
This means that, if $M_u$, $M_s$ and $\Lambda$ are fitted, $w_7$, $w_{10}$ and $w_{12}$ are automatically determined from the spin-1 meson masses by these relations. Hence, the values of these three parameters are tightly constrained by the empirical data and should be properly fitted. 

Similarly to what has been said concerning the assignments $w_6 = w_9 = 0$, a full unambiguous fitting of all the $w$'s will always require us to study their impact on the effective three-meson vertices, but we may still use this somewhat arbitrary fitting scheme at the quadratic Lagrangian level to check if the model is able to reproduce the meson spectra.

A useful systematic approach to the fitting routine may start by identifying all the conditions which involve only the $w_i$, $M_i$ and $\Lambda$; these are the 3 vector and the 3 axial vector masses, and the $f_\pi$ and $f_K$ weak decay constants. With $w_6 = w_9 = w_{13} = 0$, and with $w_1$ previously fixed, the remaining $w$'s ($w_7,w_8,w_{10},w_{11},w_{12}$), as well as $M_u,M_s$ and $\Lambda$, may be fitted using the above mentioned 8 empirical inputs. 

The fact that both constituent quark masses and the scale $\Lambda$ are fixed entirely by the spin-1 spectra and the pseudoscalar weak decay constants is a detail worthy of note. Actually, it can be shown using the mass formulae (\ref{vector-masses}) and (\ref{axial-masses}) and the explicit form of the coefficients $H_{ab}$ that 
\begin{equation}
\label{spin1-mass-formula}
\frac{2}{M_{K^*}^2 - \frac{3}{2} \left(M_u - M_s\right)^2} + \frac{2}{M_{K_1}^2 - \frac{3}{2} \left(M_u + M_s\right)^2} = \frac{1}{M_{\rho}^2} + \frac{1}{M_{\varphi}^2} + \frac{1}{M_{a_1}^2 - 6 M_u^2} + \frac{1}{M_{f_1}^2 - 6 M_s^2}.
\end{equation}
This means that, in the isospin limit ($m_u=m_d$), the model predicts a relation between $M_u$ and $M_s$ depending solely on the spin-1 meson masses. 

Furthemore, the axial-vector meson masses in (\ref{axial-masses}) may be rewritten as
\begin{equation}
\label{axial-masses-2}
M_{a_1}^2=\frac{6 M_{u}^4}{M_{u}^2-\varrho^2f_{\pi}^2}, \qquad
M_{K_1}^2=\frac{\frac{3}{2}\left(M_{u}+M_s\right)^4}{\left(M_{u}+M_s\right)^2-4\varrho^2f_K^2},
\end{equation}
i.e. in terms of constituent quark masses, the scale $\Lambda$ (through $\varrho^2$) and weak decay constants only. Together with expression (\ref{spin1-mass-formula}), these relations completely determine $M_u$, $M_s$ and $\Lambda$.

The result of the partial fitting described above may then be carried on to the remaining conditions as inputs in order to fit the rest of the parameters. The $h_i$ are already fully determined by the gap equations at this stage, so we may then focus on the 4 scalar and 4 pseudoscalar masses, the 2 mixing angle conditions and the 2 stationary phase conditions in order to fit the 11 parameters ($G,\kappa,\kappa_2,g_1,g_2,g_3,g_4,g_5,g_6,g_7,g_8$) as well as the scalar mixing angle $\theta_{\sigma}$.

From the empirical point of view, the pseudoscalar and the vector low-lying nonets are relatively well established, the former with the $\pi$, $K$, $\eta$ and $\eta'\left(958\right)$ mesons and the latter with the $\rho\left(770\right)$, $K^*\left(892\right)$, $\omega\left(782\right)$ and $\varphi\left(1020\right)$ mesons. The axial-vector nonet that we will try to fit consists of $a_1\left(1260\right)$, $K_1\left(1270\right)$, $f_1\left(1285\right)$ and $f_1\left(1420\right)$, of which $a_1\left(1260\right)$ exhibits a broader peak leading to a larger experimental mass uncertainty. Also, some authors propose $f_1\left(1510\right)$ as a member of the nonet instead of $f_1\left(1420\right)$, and there are suggestions as to $f_1\left(1285\right)$ and $f_1\left(1420\right)$ being actually $K^* \bar{K}$ molecules or tetraquark states \cite{PDG}. Meanwhile, the scalar nonet is probably the most controversial one. Models relying heavily on chiral symmetry constraints (as is the case of the model under study) usually identify the members of the low-lying scalar nonet as the $\sigma\left(500\right)$, the $\kappa \left(800\right)$, the $a_0\left(980\right)$ and the $f_0\left(980\right)$, although different approaches may establish the nonet with some other states, namely the $K^*_0\left(1430\right)$ instead of the the $\kappa \left(800\right)$ \cite{PDG}. As was done in the spin-0 version of the model, we expect that also the spin-1 extended version is able to fit this nonet.  Still concerning the scalars, the exact physical content of the corresponding measured signals is disputed, with some authors proposing significant contributions from four-quark states, gluon-balls or meson-meson molecules \cite{PDG}. The present model contemplates the admixture of four-quark components to the usual $\bar{q}q$ content of the mesons, which is arguably an advantageous feature of the approach. Yet, we are also faced with a large empirical range for the masses of the $\sigma\left(500\right)$ ($400\sim550 \text{ MeV}$) and $\kappa\left(800\right)$ ($650 \sim 850 \text{ MeV}$) mesons \cite{PDG}.

Using the empirical inputs listed in table \ref{emp-inputs}, we obtain the fitted values for the model's parameters as is shown in table \ref{fit}. 

The first noteworthy aspect of the results concerns the higher value obtained for the cut-off $\Lambda = 1633 \text{ MeV}$, which is around two times of the value fitted with the spin-0 version of the model. If we expect the model to provide meaningful results for the spin-1 meson masses, which are generally higher than those of the spin-0 ones, a higher value of this scale is desirable. In fact, the highest mass employed in the fitting ($M_{f_1} = 1426 \text{ MeV}$) is below the fitted scale value, which should be expected from an effective theory point of view. On the other hand, this value still fulfills the general requirement that $\Lambda$ is of order of chiral symmetry breaking scale $\Lambda\sim\Lambda_{\chi SB}\sim 1\,\mbox{GeV}$.

\begin{table}
\caption{The 19 input phenomenological values used in the fitting of the model parameters: the meson masses, the current quark masses and the weak decay constants (all in MeV), as well as the pseudoscalar mixing angle in degrees.\\}
\label{emp-inputs}
\begin{tabular}{ccccccccccccccccccc}
\hline
\textbf{$M_\pi$} 
&\textbf{$M_K$} 
&\textbf{$M_\eta$}
&\textbf{$M_{\eta'}$}
&\textbf{$M_\sigma$}
&\textbf{$M_\kappa$}
&\textbf{$M_{a_0}$}
&\textbf{$M_{f_0}$}
&\textbf{$M_\rho$}
&\textbf{$M_{K^*}$}
&\textbf{$M_\varphi$}
&\textbf{$M_{a_1}$}
&\textbf{$M_{K_1}$}
&\textbf{$M_{f_1}$}
&\textbf{$m_u$}
&\textbf{$m_s$}
&\textbf{$f_\pi$}
&\textbf{$f_K$}
&\textbf{$\theta_\phi$}\\ 
\hline
  138
& 496 
& 548 
& 958 
& 500 
& 850 
& 980 
& 980 
& 778 
& 893 
& 1019 
& 1270 
& 1274 
& 1426 
& 4 
& 100 
& 92 
& 111 
& -15$^{\circ}$ \\
\hline
\end{tabular}
\end{table}

\begin{table}
\caption{The 24 model parameters obtained as the results of the fit. The value of $w_1$ shown here has been externally fixed, with $w_6 = w_9 = w_{13} = 0$. \\}
\label{fit}
\begin{tabular}{cccccccccccc}
\hline
  \textbf{$G$} 
& \textbf{$\kappa$} 
& \textbf{$g_1$} 
& \textbf{$g_2$} 
& \textbf{$\kappa_2$} 
& \textbf{$g_3$} 
& \textbf{$g_4$} 
& \textbf{$g_5$} 
& \textbf{$g_6$} 
& \textbf{$g_7$} 
& \textbf{$g_8$} 
& \textbf{$\theta_\sigma$} \\
  $\text{GeV}^{-2}$
& $\text{GeV}^{-5}$
& $\text{GeV}^{-8}$
& $\text{GeV}^{-8}$
& $\text{GeV}^{-3}$
& $\text{GeV}^{-6}$
& $\text{GeV}^{-6}$
& $\text{GeV}^{-4}$
& $\text{GeV}^{-4}$
& $\text{GeV}^{-4}$
& $\text{GeV}^{-4}$
& $\text{degrees}$ \\
\hline 
  2.54 
& -2.66 
& 15.3 
& -35.2 
& 0.143 
& -148 
& 36.1 
& -21.9 
& -115 
& -32.6 
& -21.8
& 25.1$^{\circ}$ \\
\hline
\hline
  \textbf{$\Lambda$}
& \textbf{$M_u$}
& \textbf{$M_s$}
& \textbf{$w_1$} 
& \textbf{$w_6$} 
& \textbf{$w_7$} 
& \textbf{$w_8$} 
& \textbf{$w_9$} 
& \textbf{$w_{10}$} 
& \textbf{$w_{11}$} 
& \textbf{$w_{12}$}
& \textbf{$w_{13}$}\\
  $\text{MeV}$
& $\text{MeV}$
& $\text{MeV}$
& $\text{GeV}^{-2}$
&
& $\text{GeV}^{-8}$
& $\text{GeV}^{-8}$
& 
& $\text{GeV}^{-6}$
& $\text{GeV}^{-6}$
& $\text{GeV}^{-4}$
&  \\
\hline
  1633
& 244
& 508
& -10 
& 0
& -1903 
& 2505 
& 0
& -2540 
& 1425 
& -1523
& 0 \\ 
\hline
\end{tabular}
\end{table}

Constituent quark masses are fitted to lower values than those of the spin-0 model, with a more significant difference in $M_u$. Nonetheless, these are still within reasonable values, and the mass difference $M_s - M_u$ is enhanced. Together with $\Lambda$, $M_u$ and $M_s$ form a set of three parameters which may be fixed solely on the basis of the spin-1 spectra and the pseudoscalar weak decay constants as shown in expressions (\ref{spin1-mass-formula}) and (\ref{axial-masses-2}). This means that the experimental uncertainties of the scalar sector do not affect these results, leaving little room for variation of the values of $M_u$, $M_s$ and $\Lambda$ within the relatively well-defined experimental ranges of the spin-1 meson masses and the pion and kaon weak decay constants. Variations within said ranges consistently yield similarly larger values for $\Lambda$ and smaller values for $M_u$ and $M_s$ than those found in the spin-0 version of the model. The scalar mixing angle fitted value $\theta_\sigma = 25.1^{\circ}$ is also well within the (model dependent) empirical range.

The $w$'s (disregarding those which are externally fixed or set to $0$) are fitted to values of the order of $\mathcal{O}\left(10^3\right)$ in their respective units, yielding significant contributions to the spin-0 sectors through the $V\sigma$- and $A\phi$-mixing mechanisms. This in turn affects the old parameters' fitted values, which all turn out considerably smaller (in absolute value) than they were in the spin-0 version of the model. If we compare old and new parameters' values of the same dimensionality, we identify a consistent proportion of 1 to 2 orders of magnitude between them. Different prescriptions for the external fixing of $w_1$ and $w_{13}$ have little to no impact in this fact. This may be regarded as a quantitative statement about relative weights of spin-0 and spin-1 multi-quark vertices in the effective description of the dynamics of strongly interacting particles, in support of the importance of including spin-1 modes in the model.

The spin-0 spectra is reproduced similarly to what has been done in \cite{model1,model2} with the spin-0 version of the model. The differences arise through the contributions of the spin-1 $w_i$ parameters and the fact that the scalar mixing angle $\theta_\sigma$ is fitted together with the effective couplings, yielding a value consistent with those provided in the literature \cite{Gokalp2005,Aliev2010}. This fitted value is, of course, subject to variations due to the large uncertainties in the empirical masses of the $\kappa\left(800\right)$ and $\sigma\left(500\right)$ mesons. This dependency is illustrated in Table \ref{scalar-table}, where the value of the scalar mixing angle is shown for different combinations of the above mentioned masses (all other input being as given in Table \ref{emp-inputs}). We can see a very significant variation of $\theta_\sigma$ with the $\kappa\left(800\right)$ mass, which seems to require this mass to be on the higher side of its empirical range for $\theta_\sigma$ to be within reasonable values.

\begin{table}
\caption{Values of the scalar mixing angle for different prescriptions of $M_\kappa$ and $M_\sigma$ (with all other empirical inputs as in table \ref{emp-inputs}).\\}
\label{scalar-table}
\begin{tabular}{cccccccc}
\hline
  $M_{\kappa}$ 
&
& \multicolumn{2}{c}{750}
& \multicolumn{2}{c}{800}
& \multicolumn{2}{c}{850} \\
\hline
 $M_{\sigma}$
&
& 400 
& 500 
& 400 
& 500 
& 400 
& 500 \\
  $\theta_\sigma$
& 
& 42.9$^\circ$ 
& 44.9$^\circ$ 
& 35.7$^\circ$ 
& 36.9$^\circ$ 
& 25.5$^\circ$ 
& 25.1$^\circ$ \\
\hline
\end{tabular}
\end{table}

The high number of effective couplings which the model introduces may be regarded as a shortcoming, based on the notion that a sufficiently high number of parameters is a sufficient condition to fit any kind of data, making the modelling rather arbitrary and devoid of physical meaning. However, the way the effective couplings are introduced in the model is not at all arbitrary, lending themselves to strict symmetry constraints which are provided by the underlying fundamental physics at work. Furthermore, we should take a look at the way the $w_i$'s enter the model's expressions for the observables considered in this study to realize how these symmetry constraints strongly bind the parameters' eventual arbitrariness. From the 13 $w_i$'s which are introduced through the effective multi-quark vertices, only 9 appear in the quadratic part of the Lagrangian; from these 9, not all are really independent, with e.g. $w_1$, $w_6$ and $w_9$ clustering into a single effective contribution. These somewhat subtle and intricate relations effectively express the symmetry constraints of the model and, hence, the underlying physics. The ability of the model to reproduce the low-lying meson spectra should not be taken as an \textit{a priori} feature of a large number of parameters, but rather as a successful capturing of relevant physical content. 

\section{Coupling constants in natural units 
\label{s6}}

The above discussion would be incomplete without giving some qualitative arguments based on naive dimensional analysis applied to the effective multi-quark Lagrangian. Although these arguments cannot be trusted to any great numerical accuracy, they provide a qualitative guide to the presented picture of the family of multi-quark couplings and interactions. Our guiding principle in this consideration is the idea of naturalness (in the Dirac sense), according to which after extracting the dimensional scales from a term of the Lagrangian, the remaining dimensionless coefficient should be of order of unity. Naively, we did this in eqs. (\ref{main0})-(\ref{main1}). Here we would like to make our consideration more detailed. 

Indeed, in the problem considered we have several important scales. First, of course, a dimensionful parameter $\Lambda = 1.633\ \mbox{GeV}$ which estimates the chiral symmetry breaking scale and suppresses non-renormalizable terms in an effective multi-quark Lagrangian. However, we might also want to consistently count powers of the effective constituent quark mass $M=244\ \mbox{MeV}$ (we will neglect in our naive analysis the difference between strange and non strange quark masses) and the pion weak decay constant $f_\pi=92\ \mbox{MeV}$. The mass $M$ is a characteristic of chirality violation at the vertex. For instance, if the Lagrangian contains the quark bilinears $\bar q_Rq_L$ or $\bar q_Lq_R$, then such vertex changes chirality by the value $|\Delta\chi|=2$. Our Lagrangian includes these transitions through the terms (\ref{sb}) and (\ref{asb}). On the other hand, the quark bilinears $\bar q_L\gamma_\mu q_L$ or $\bar q_R\gamma_\mu q_R$ do not change chirality. Thus, for them $|\Delta\chi|=0$, and this is true for the vector combinations (\ref{rv}) and (\ref{lv}). 
   
The pion decay constant $f_\pi$ is a dimensionful parameter which governs the dynamics of the Goldstone boson fields. At low energy it is small compared to $\Lambda$ and naturally appears when one bosonizes the multi-quark interactions. Taking all these scales into account, we come to the formula 
\begin{equation}
\label{da}
\mathcal{L}=\bar{c}\left(\frac{\pi}{f}\right)^A\left(\frac{q}{f\sqrt{\Lambda}}\right)^B\left(\frac{M}{\Lambda}\right)^C\left(\frac{\partial}{\Lambda}\right)^D\left(\frac{\chi}{M}\right)^E f^2\Lambda^2,
\end{equation}
where $\bar{c}$ is the dimensionless constant of order of unity (for natural units), $A$ is the meson field power, $B$ is the quark field power, $C=|\Delta\chi|/2$ describes the chirality violation at the quark part of the vertex, $D$ is the number of derivatives, and $E$ counts the explicit symmetry breaking effects induced by the external field $\chi\sim m$. Comparing our result with the one of Manohar and Georgi \cite{Georgi84}, it should be noted that our Lagrangian at the quark level does not have derivative interactions. However, if one would like to analyse the couplings of the effective meson Lagrangian which results from bosonization, one should include this term too in accordance with \cite{Georgi84}. On the other hand, they do not consider explicit symmetry breaking effects and, as a result, they do not have the term with $\chi$, as we have.

The prescription (\ref{da}) produces a set of coefficients for the higher-dimension operators which are consistent with naive dimensional analysis. Indeed, to have a feeling that this prescription agrees with our naive expectations, we show the order of the following terms:
\begin{eqnarray}
& \pi\bar q\gamma_5q \to \pi\bar q\gamma_5q\left(\frac{1}{f}\right)\left(\frac{1}{f\sqrt{\Lambda}}\right)^2\left(\frac{M}{\Lambda}\right)f^2\Lambda^2 =\left(\frac{M}{f}\right)\pi\bar q\gamma_5q.\\
& \bar q\gamma_\mu\partial^\mu q\to \bar q\gamma_\mu\partial^\mu q\left(\frac{1}{f\sqrt{\Lambda}}\right)^2\left(\frac{1}{\Lambda}\right)f^2\Lambda^2=\bar q\gamma_\mu\partial^\mu q. \\
& \bar qmq \to \bar qmq \left(\frac{1}{f\sqrt{\Lambda}}\right)^2\left(\frac{M}{\Lambda}\right)\left(\frac{1}{M}\right) f^2\Lambda^2 =\bar qmq. 
\end{eqnarray}
We see that dimensional arguments work well for the kinetic and mass terms of the quark Lagrangian, and it even gives a correct estimate for the coupling of the pion to two quarks, which is $g_\pi=M/f$ according to a quark analogue of the Goldberger-Treiman relation.

It follows then that the scaling factor $S$ for the conversion from dimensionful $c$ to natural $\bar c$ coupling constants is
\begin{equation}
\bar c = Sc, \qquad S=\frac{f^{A+B-2}\Lambda^{\frac{B}{2}+C+D-2}}{M^{C-E}}.
\end{equation}  
Using this result, one can obtain the dimensionless values of the coupling constants of the Lagrangian (\ref{main0})-(\ref{main1}). We collect them in Table \ref{nat-units}. It is enough to know the order of the corresponding values. 

\begin{table}
\caption{The order of model parameters in natural units. We collect the non-zero dimensional couplings $c$, scaling factors $S$ and the order of dimensionless coupling constants $\bar c=Sc$.\\}
\label{nat-units}
\begin{tabular}{cccccccccccccccccc}
\hline
  \text{c}
& \textbf{$G$} 
& \textbf{$\kappa$} 
& \textbf{$g_1$} 
& \textbf{$g_2$} 
& \textbf{$\kappa_2$} 
& \textbf{$g_3$} 
& \textbf{$g_4$} 
& \textbf{$g_5$} 
& \textbf{$g_6$} 
& \textbf{$g_7$} 
& \textbf{$g_8$} 
& \textbf{$w_1$} 
& \textbf{$w_7$} 
& \textbf{$w_8$} 
& \textbf{$w_{10}$} 
& \textbf{$w_{11}$} 
& \textbf{$w_{12}$} \\ 
  \hline
  \text{S}
& $\frac{f^2\Lambda^2}{M^2}$
& $\frac{f^4\Lambda^4}{M^3}$
& $\frac{f^6\Lambda^6}{M^4}$
& $\frac{f^6\Lambda^6}{M^4}$
& $\frac{f^2\Lambda^2}{M}$
& $\frac{f^4\Lambda^4}{M^2}$
& $\frac{f^4\Lambda^4}{M^2}$
& $f^2\Lambda^2$
& $f^2\Lambda^2$
& $f^2\Lambda^2$
& $f^2\Lambda^2$ 
& $f^2$ 
& $\frac{f^6\Lambda^4}{M^2}$
& $\frac{f^6\Lambda^4}{M^2}$
& $f^4\Lambda^2$
& $f^4\Lambda^2$
& $f^2M^2$ \\
  \text{$\bar c$}
&  1.0
& -0.1 
&  0.05
& -0.1 
&  0.01 
& -1.3 
&  0.3 
& -0.5
& -2.6
& -0.7 
& -0.5 
& -0.1 
& -0.1 
&  0.1 
& -0.5 
&  0.3 
& -0.8 \\ 
\hline
\end{tabular}
\end{table}

Let us discuss the naturalness of these values. The couplings $\bar G, \bar g_3, \bar g_5, \bar g_6, \bar g_7, \bar g_8, \bar w_{10}, \bar w_{12}$ are of order 1 and therefore they are natural. The couplings $\bar\kappa, \bar g_1, \bar g_2, \bar g_4, \bar w_1, \bar w_7, \bar w_8, \bar w_{11}$ are one order suppressed. The coupling $\bar\kappa_2$ is two orders less than the main set. They are unnatural. There are several reasons for them to be small. For example, $\bar\kappa$ and $\bar\kappa_2$ both break explicitly the axial $U_A(1)$ symmetry, violate Zweig's rule, and $\bar\kappa_2$ additionally breaks chiral symmetry explicitly. The eight quark interactions with couplings $\bar g_1, \bar g_2, \bar w_7, \bar w_8$ are $1/N_c$ suppressed compared to four-quark interactions. The couplings $\bar g_4$ and $\bar w_{11}$ break chiral symmetry explicitly. The relatively small value of the coupling $\bar w_1$ of four-quark vector interactions is a bit surprising. However, it is known that vector excitations need more energy to be generated. For instance, in chiral perturbation theory they appear only at $p^6$ order. All these naturalness considerations follow 't Hooft's notion of naturalness: that a parameter is naturally small if setting it to zero enhances the 
symmetry of the theory.

\section{Conclusions and outlook 
\label{s7}}

We studied a generalized three-flavor NJL-type model with spin-1 mesons included. As a new aspect, we have considered the explicit symmetry breaking (ESB) effects induced by the multi-quark interactions. The latter are supposed to appear at low energies as a result of long scale QCD dynamics. The standard quark mass term $\bar qmq$ is considered to be a leading order term in the hierarchy of possible multi-quark interactions. These effects are known to be important in chiral perturbation theory (due to a large strange quark mass $m_s\sim 100\,\mbox{MeV}$). The effective model with multi-quark interactions naturally incorporates the vertices with higher powers of current quark masses, in terms of which the problem can easily be formulated. We collected all such effective interactions (without derivatives) and investigated their influence on the mass spectrum of spin-0 and spin-1 mesons. Our result shows that the next to leading order current-quark-mass corrections are tractable and essentially improve our description of meson spectra. This is the main result of our work. 

One should note that we are still far away from a satisfactory theory for collective quark states. The approach considered here adds to the many known attempts in this direction an interesting new feature - the possibility to study directly the internal mechanism of the bound states' formation which includes not only the leading effect of quark - antiquark pairing but also takes into account the subleading effects due to the admixture of $\bar{q}q\bar{q}q$ components and ESB.      

Our analysis can be extended in several directions. First, the large amount of phenomenological results give us the hope that we may estimate the importance of explicit symmetry breaking phenomena for some processes. We are working in this direction. Second, nowadays it is getting clear that the multi-quark interactions can be important for the description of quark matter in a strong magnetic background (for instance, in stars). It would be interesting to understand which set of the
effective quark-mass dependent interactions is of importance here. A further motivation comes from the hadronic matter studies in a hot and dense environment. The critical points of the phase diagram and even the type of phase transitions are sensitive to the quark masses.

\appendix
\section{The expressions for three-index coefficients $h_{abc}$ and $H_{abc}$
\label{app1}}

The stationary phase equations of motion (\ref{SPA-conds}) fix all higher order coefficients of the series (\ref{SPA-exp}). Here we show the result of our calculations of three-index coefficients $h_{abc}^{(i)}$ and $H_{abc}^{(i)}$. Three of them ($h_{abc}^{(1,2,3)}$) have been already computed. We present them here for completeness:    
\begin{align}
h_{fgh}^{\left(1\right)} & = h_{af}^{\left(1\right)} h_{bg}^{\left(1\right)} h_{ch}^{\left(1\right)} \left[ \frac{3\kappa}{4} A_{abc} + g_1 \left(2 \delta_{ab} h_c + \delta_{bc} h_a\right) + \frac{g_2}{2} \left(2d_{abe}d_{cde} + d_{ade}d_{bce}\right) h_d \right. \nonumber \\
& \left. + \frac{g_3}{4} \left(3d_{abe}d_{cde} - f_{abe}f_{cde}\right) m_d + \frac{g_4}{2} \left(2 \delta_{ab} m_c + \delta_{bc} m_a\right) \right], \\
h_{fgh}^{\left(2\right)} & = h_{af}^{\left(1\right)} h_{bg}^{\left(2\right)} h_{ch}^{\left(2\right)} \left[ -\frac{3\kappa}{4} A_{abc} + g_1 \delta_{bc} h_a + \frac{g_2}{2} \left(d_{ade}d_{bce} - 2f_{abe}f_{cde}\right) h_d \right. \nonumber \\
& \left. - \frac{g_3}{4} \left(3f_{abe}f_{cde} - d_{abe}d_{cde}\right) m_d + \frac{g_4}{2} \delta_{bc} m_a \right], \\
h_{fgh}^{\left(3\right)} & = h_{af}^{\left(2\right)} h_{bg}^{\left(2\right)} h_{ch}^{\left(1\right)} \left[ -\frac{3\kappa}{2} A_{abc} + 2g_1 \delta_{ab} h_c + g_2 \left(d_{abe}d_{cde} + f_{ace}f_{bde} + f_{ade}f_{bce}\right) h_d \right. \nonumber \\
& \left. + \frac{g_3}{2} \left(d_{abe}d_{cde} + f_{ace}f_{bde} + f_{ade}f_{bce}\right) m_d + g_4 \delta_{ab} m_c \right].
\end{align}

\noindent Other seven coefficients $H_{abc}^{(1-7)}$ are new. They are
\begin{align}
H_{fgh}^{\left(1\right)} & = h_{af}^{\left(1\right)} H_{bg}^{\left(1\right)} H_{ch}^{\left(1\right)} \left[ w_6 \delta_{bc} h_a + \frac{w_7}{4} \left(d_{ace}d_{bde} + f_{ace}f_{bde}\right) h_d + \frac{w_8}{2} d_{ade}d_{bce} h_d \right. \nonumber \\
& \left. + w_9 \delta_{bc} m_a + \frac{w_{10}}{4} \left(d_{abe}d_{cde} + f_{abe}f_{cde}\right) m_d + \frac{w_{11}}{2} d_{ade}d_{bce} m_d \right], \\
H_{fgh}^{\left(2\right)} & = h_{af}^{\left(1\right)} H_{bg}^{\left(2\right)} H_{ch}^{\left(2\right)} \left[ w_6 \delta_{bc} h_a - \frac{w_7}{4} \left(d_{ace}d_{bde} + f_{ace}f_{bde}\right) h_d + \frac{w_8}{2} d_{ade}d_{bce} h_d \right. \nonumber \\
& \left. + w_9 \delta_{bc} m_a - \frac{w_{10}}{4} \left(d_{abe}d_{cde} + f_{abe}f_{cde}\right) m_d + \frac{w_{11}}{2} d_{ade}d_{bce} m_d \right], \\
H_{fgh}^{\left(3\right)} & = h_{af}^{\left(2\right)} H_{bg}^{\left(1\right)} H_{ch}^{\left(2\right)} \left[ \frac{w_7}{2} \left(f_{abe}d_{cde} - d_{abe}f_{cde}\right) h_d + w_8 f_{ade}d_{bce} h_d \right. \nonumber \\
& \left. + \frac{w_{10}}{2} \left(d_{ace}f_{bde} - f_{ace}d_{bde}\right) m_d + w_{11} f_{ade}d_{bce} m_d \right],\\
H_{fgh}^{\left(4\right)} & = H_{af}^{\left(1\right)} h_{bg}^{\left(1\right)} H_{ch}^{\left(1\right)} \left[ 2w_6 \delta_{ac} h_b + \frac{w_7}{2} \left(d_{abe}d_{cde} - f_{abe}f_{cde}\right) h_d + w_8 d_{ace}d_{bde} h_d \right. \nonumber \\
& \left. + 2w_9 \delta_{ac} m_b + \frac{w_{10}}{2} \left(d_{abe}d_{cde} - f_{abe}f_{cde}\right) m_d + w_{11} d_{ace}d_{bde} m_d \right], \\
H_{fgh}^{\left(5\right)} & = H_{af}^{\left(1\right)} h_{bg}^{\left(2\right)} H_{ch}^{\left(2\right)} \left[ - \frac{w_7}{2} \left(d_{abe}f_{cde} + f_{abe}d_{cde}\right) h_d + w_8 d_{ace}f_{bde} h_d \delta_{ac} m_b \right. \nonumber \\
& \left.  + \frac{w_{10}}{2} \left(f_{ade}d_{bce} - d_{ade}f_{bce}\right) m_d + w_{11} d_{ace}f_{bde} m_d \right], \\
\end{align}

\begin{align}
H_{fgh}^{\left(6\right)} & = H_{af}^{\left(2\right)} h_{bg}^{\left(2\right)} H_{ch}^{\left(1\right)} \left[ - \frac{w_7}{2} \left(f_{ade}d_{bce} - d_{ade}f_{bce}\right) h_d + w_8 d_{ace}f_{bde} h_d \right. \nonumber \\
& \left.  + \frac{w_{10}}{2} \left(f_{abe}d_{cde} + d_{abe}f_{cde}\right) m_d + w_{11} d_{ace}f_{bde} m_d \right], \\
H_{fgh}^{\left(7\right)} & = H_{af}^{\left(2\right)} h_{bg}^{\left(1\right)} H_{ch}^{\left(2\right)} \left[ 2w_6 \delta_{ac} h_b - \frac{w_7}{2} \left(d_{abe}d_{cde} - f_{abe}f_{cde}\right) h_d + w_8 d_{ace}d_{bde} h_d \right. \nonumber \\
& \left. + 2w_9 \delta_{ac} m_b - \frac{w_{10}}{2} \left(d_{abe}d_{cde} - f_{abe}f_{cde}\right) m_d + w_{11} d_{ace}d_{bde} m_d \right].
\end{align}

\section{Basic steps in diagonalizing $V\sigma$- and $A\phi$-mixing 
\label{app2}}

For the sake of completeness we describe in this appendix the basic steps made to avoid the $V\sigma$- and $A\phi$-mixing from the meson Lagrangian. As in the main text, we focus only on the quadratic part of the Lagrangian density, where we preform the following shifts: 
\begin{eqnarray*}
&V_{a\mu}\to V_{a\mu} + 2k_a f_{abc} M_b \partial_{\mu} \sigma_c \\ 
&A_{a\mu} \to A_{a\mu} + 2k'_a d_{abc} M_b \partial_{\mu} \phi_c,
\end{eqnarray*}
as well as $F_{a\mu\nu}^{\left(V\right)} \to F_{a\mu\nu}^{\left(V\right)}$ and $F_{a\mu\nu}^{\left(A\right)} \to F_{a\mu\nu}^{\left(A\right)}$. The quadratic part of the Lagrangian now reads
\begin{align}
\label{quad-Lag-2}
\mathcal{L}_{bos}^{\left(2\right)} & = \frac{1}{2} \left(h_{ab}^{\left(1\right)} \sigma_a \sigma_b + h_{ab}^{\left(2\right)} \phi_a \phi_b + H_{ab}^{\left(1\right)} V^{\mu}_a V_{b\mu} + H_{ab}^{\left(2\right)} A^{\mu}_a A_{b\mu} \right) \nonumber \\
& + 2 \left( H_{ab}^{\left(1\right)} V^{\mu}_a k_b f_{bcd} M_c \partial_{\mu} \sigma_d + H_{ab}^{\left(2\right)} A^{\mu}_a k'_b d_{bcd} M_c \partial_{\mu} \phi_d \right. \nonumber \\
& \left. + H_{ab}^{\left(1\right)} f_{acd} f_{bc'd'} k_a k_b M_c M_{c'} \partial^{\mu} \sigma_d \partial_{\mu} \sigma_{d'} + H_{ab}^{\left(2\right)} d_{acd} d_{bc'd'} k'_a k'_b M_c M_{c'} \partial^{\mu} \phi_d \partial_{\mu} \phi_{d'} \right) \nonumber \\
& + \frac{N_c I_1}{16\pi^2} \text{tr}_F \left\lbrace \left(\partial^{\mu} \sigma \right)\left(\partial_{\mu} \sigma \right) + \left(\partial^{\mu} \phi \right)\left(\partial_{\mu} \phi \right) - \frac{1}{3} \left( F^{\mu\nu}_{\left(V\right)} F_{\mu\nu}^{\left(V\right)} + F^{\mu\nu}_{\left(A\right)} F_{\mu\nu}^{\left(A\right)} \right) + \frac{2I_0}{I_1} \left( \sigma^2 + \phi^2 \right) \right. \nonumber \\
& \left. - \left\lbrace \sigma, M \right\rbrace^2 + \left[\phi, M \right]^2 - \left( \Delta_{ud} \lambda_3  + \frac{\Delta_{us} + \Delta_{ds}}{\sqrt{3}} \lambda_8 \right) \left(\sigma^2 + \phi^2\right) \right\rbrace \nonumber \\
+ & \frac{N_c I_1}{2\pi^2} \left[ f_{ace}f_{bde} \left(V^{\mu}_a + 2k_a f_{amn} M_m \partial^{\mu} \sigma_n\right) \left(V_{b\mu} + 2k_b f_{bm'n'} M_{m'} \partial_{\mu} \sigma_{n'}\right) M_c M_d \right. \nonumber \\
& + d_{ace}d_{bde} \left(A^{\mu}_a + 2k'_a d_{amn} M_m \partial^{\mu} \phi_n\right) \left(A_{b\mu} + 2k'_b d_{bm'n'} M_{m'} \partial_{\mu} \phi_{n'}\right) M_c M_d \nonumber \\
& \left. + f_{abc} \left( V^{\mu}_a + 2 k_a f_{ade} M_d \partial^{\mu} \sigma_e \right) \partial_{\mu} \sigma_b M_c - d_{abc} \left( A^{\mu}_a + 2 k'_a d_{ade} M_d \partial^{\mu} \phi_e \right) \partial_{\mu} \phi_b M_c \right].
\end{align}

\noindent We may then collect the mixing terms and write them as
\begin{align}
\label{mixing-terms}
\mathcal{L}_{V\sigma}^{\left(2\right)} & = \frac{N_c I_1}{2\pi^2} V^{\mu}_a \partial_{\mu} \sigma_b M_c \left[ 4k_d f_{cbd} \left( \frac{\pi^2 H_{ad}^{\left(1\right)}}{N_c I_1} + f_{ame} f_{dne} M_m M_n  \right) + f_{abc} \right] \nonumber \\
\mathcal{L}_{A\phi}^{\left(2\right)} & = \frac{N_c I_1}{2\pi^2} A^{\mu}_a \partial_{\mu} \phi_b M_c \left[ 4k'_d d_{cbd} \left( \frac{\pi^2 H_{ad}^{\left(2\right)}}{N_c I_1} + d_{ame} d_{dne} M_m M_n  \right) - d_{abc} \right].
\end{align}
The diagonalization of the Lagrangian requires the vanishing of the coefficient for each combination of $V\sigma$ or $A\phi$ field components in (\ref{mixing-terms}). These diagonalization conditions may be written as
\begin{align}
\label{diag-conds}
M_ck_df_{bcd}\left[ \frac{\pi^2 H_{ad}^{\left(1\right)}}{N_c I_1} +f_{ame}f_{dne}M_m M_n \right] & = \frac{1}{4}M_cf_{abc} \nonumber \\
M_c k'_d d_{bcd} \left[ \frac{\pi^2 H_{ad}^{\left(2\right)}}{N_c I_1} + d_{ame} d_{dne} M_m M_n \right] & = \frac{1}{4}M_c d_{abc},
\end{align}
which must be obeyed for each combination of $a,b$ indices in the range $\left\lbrace 0,1,\dots,8 \right\rbrace$. These relations impose conditions on the constants $k_a$ and $k'_a$, and enable us to get rid of mixing. As a result we come to expressions (\ref{k}) and (\ref{k'}) of the main text.

\section{Mass Lagrangians 
\label{app3}}

Following the introduction of meson fields and mixing angles in Section \ref{s4}, we may write down the mass terms from the Lagrangian densities of the spin-0 fields as
\begin{align}
\label{quad-Lag-sigma-final}
\mathcal{L}_{\sigma}^{\left(mass\right)} & = \frac{1}{2} \left(a_0^0\right)^2 \left[ \varrho^2 \left( 2h_{uu}^{\left(1\right)} - 2h_{ud}^{\left(1\right)} - \frac{h_u}{M_u} \right) - 4 M_u^2 \right] \nonumber \\
& + \frac{1}{2} f_0^2 \left\lbrace \cos^2{\psi_{\sigma}} \left[ \varrho^2 \left( 2h_{uu}^{\left(1\right)} - 2h_{ud}^{\left(1\right)} - \frac{h_u}{M_u} \right) - 4 M_u^2 \right] + \sin^2{\psi_{\sigma}} \left[ \varrho^2 \left( 2h_{ss}^{\left(1\right)} - \frac{h_s}{M_s} \right) - 4 M_s^2 \right] \right. \nonumber \\
& \left. - 4 \sqrt{2} \sin{\psi_{\sigma}}\cos{\psi_{\sigma}} \varrho^2 h_{us}^{\left(1\right)} \right\rbrace \nonumber \\
& + \frac{1}{2} \sigma^2 \left\lbrace \sin^2{\psi_{\sigma}} \left[ \varrho^2 \left( 2h_{uu}^{\left(1\right)} - 2h_{ud}^{\left(1\right)} - \frac{h_u}{M_u} \right) - 4 M_u^2 \right] + \cos^2{\psi_{\phi}} \left[ \varrho^2 \left( 2h_{ss}^{\left(1\right)} - \frac{h_s}{M_s} \right) - 4 M_s^2 \right] \right. \nonumber \\
& \left. + 4 \sqrt{2} \sin{\psi_{\sigma}}\cos{\psi_{\sigma}} \varrho^2 h_{us}^{\left(1\right)} \right\rbrace \nonumber \\
& + f_0\sigma \left( \sin{\psi_{\sigma}}\cos{\psi_{\sigma}} \left\lbrace \left[ \varrho^2 \left( 2h_{uu}^{\left(1\right)} - 2h_{ud}^{\left(1\right)} - \frac{h_u}{M_u} \right) - 4 M_u^2 \right] - \left[ \varrho^2 \left( 2h_{ss}^{\left(1\right)} - \frac{h_s}{M_s} \right) - 4 M_s^2 \right] \right\rbrace \right. \nonumber \\
& \left. + 2\sqrt{2}\left( \cos^2{\psi_{\sigma}} - \sin^2{\psi_{\sigma}} \right) \varrho^2 h_{us}^{\left(1\right)} \right) 
+ a_0^{+} a_0^{-} \left[ \varrho^2 \left( h_{11}^{\left(1\right)} - \frac{h_u}{M_u} \right) - 4 M_u^2 \right] \nonumber \\
& + \left(\kappa^{+}\kappa^{-} + \kappa^0\bar{\kappa}^0\right) \left( \varrho^2 + \frac{\left( M_u - M_s \right)^2}{H_{44}^{\left(1\right)}} \right) \left( h_{44}^{\left(1\right)} - \frac{h_u - h_s}{M_u - M_s} \right).
\end{align}
\begin{align}
\label{quad-Lag-phi-final}
\mathcal{L}_{\phi}^{\left(mass\right)} & = \frac{1}{2} \left(\pi^0\right)^2 \left( \varrho^2 + \frac{2 M_u^2}{H_{uu}^{\left(2\right)}} \right) \left( 2h_{uu}^{\left(2\right)} - 2h_{ud}^{\left(2\right)} - \frac{h_u}{M_u} \right) \nonumber \\
& + \frac{1}{2} \eta^2 \left[ \cos^2{\psi_{\phi}} \left( \varrho^2 + \frac{2 M_u^2}{H_{uu}^{\left(2\right)}} \right) \left( 2h_{uu}^{\left(2\right)} + 2h_{ud}^{\left(2\right)} - \frac{h_u}{M_u} \right) + \sin^2{\psi_{\phi}} \left( \varrho^2 + \frac{2 M_s^2}{H_{ss}^{\left(2\right)}} \right) \left( 2h_{ss}^{\left(2\right)} - \frac{h_s}{M_s} \right) \right. \nonumber \\
& \left. - 4 \sin{\psi_{\phi}}\cos{\psi_{\phi}} \sqrt{2\left( \varrho^2 + \frac{2 M_u^2}{H_{uu}^{\left(2\right)}} \right) \left( \varrho^2 + \frac{2 M_s^2}{H_{ss}^{\left(2\right)}} \right)} h_{us}^{\left(2\right)} \right] \nonumber \\
& + \frac{1}{2} \eta'^2 \left[ \sin^2{\psi_{\phi}} \left( \varrho^2 + \frac{2 M_u^2}{H_{uu,dd}^{\left(2\right)}} \right) \left( 2h_{uu}^{\left(2\right)} + 2h_{ud}^{\left(2\right)} - \frac{h_u}{M_u} \right) + \cos^2{\psi_{\phi}} \left( \varrho^2 + \frac{2 M_{s}^2}{H_{ss}^{\left(2\right)}} \right) \left( 2h_{ss}^{\left(2\right)} - \frac{h_s}{M_s} \right) \right. \nonumber \\
& \left. + 4 \sin{\psi_{\phi}}\cos{\psi_{\phi}} \sqrt{2\left( \varrho^2 + \frac{2 M_u^2}{H_{uu}^{\left(2\right)}} \right) \left( \varrho^2 + \frac{2 M_s^2}{H_{ss}^{\left(2\right)}} \right)} h_{us}^{\left(2\right)} \right] \nonumber \\
& + \eta\eta' \left[ \sin{\psi_{\phi}}\cos{\psi_{\phi}} \left( \left( \varrho^2 + \frac{2 M_u^2}{H_{uu}^{\left(2\right)}} \right) \left( 2h_{uu}^{\left(2\right)} + 2h_{ud}^{\left(2\right)} - \frac{h_u}{M_u} \right) - \left( \varrho^2 + \frac{2 M_{s}^2}{H_{ss}^{\left(2\right)}} \right) \left( 2h_{ss}^{\left(2\right)} - \frac{h_s}{M_s} \right) \right) \right. \nonumber \\
& \left. + 2\left( \cos^2{\psi_{\phi}} - \sin^2{\psi_{\phi}} \right) \sqrt{2\left( \varrho^2 + \frac{2 M_u^2}{H_{uu}^{\left(2\right)}} \right) \left( \varrho^2 + \frac{2 M_s^2}{H_{ss}^{\left(2\right)}} \right)} h_{us}^{\left(2\right)} \right] \nonumber \\
& + \pi^{+}\pi^{-} \left( \varrho^2 + \frac{4 M_u^2}{H_{11}^{\left(2\right)}} \right) \left( h_{11}^{\left(2\right)} - \frac{h_u}{M_u} \right) \nonumber \\
& + \left(K^{+}K^{-} + K^0\bar{K}^0\right) \left( \varrho^2 + \frac{\left( M_u + M_s \right)^2}{H_{44}^{\left(2\right)}} \right) \left( h_{44}^{\left(2\right)} - \frac{h_u + h_s}{M_u + M_s} \right).
\end{align}
The only remaining mixing terms in (\ref{quad-Lag-sigma-final}) and (\ref{quad-Lag-phi-final}) yield diagonalization conditions in terms of the mixing angles. These can be written in the forms
\begin{align}
\label{angle-conds}
\tan{2\psi_{\sigma}} & = \frac{4 \sqrt{2} \varrho^2 h_{us}^{\left(1\right)}}{\left[\varrho^2 \left(\frac{h_{u}}{M_{u}} - 2 h_{uu}^{\left(1\right)} - 2 h_{ud}^{\left(1\right)}\right) + 4 M_{u}^2\right] - \left[\varrho^2 \left(\frac{h_s}{M_s}- 2 h_{ss}^{\left(1\right)}\right) + 4 M_s^2\right]} \nonumber \\
\tan{2\psi_{\phi}} & = \frac{4 \sqrt{2\left( \varrho^2 + \frac{2 M_{u}^2}{H_{uu}^{\left(2\right)}} \right) \left( \varrho^2 + \frac{2 M_s^2}{H_{ss}^{\left(2\right)}} \right)} h_{us}^{\left(2\right)}}{\left(\varrho^2 + \frac{2 M_{u,d}^2}{H_{uu}^{\left(2\right)}}\right) \left(\frac{h_{u}}{M_{u}} - 2 h_{uu}^{\left(2\right)} - 2 h_{ud}^{\left(2\right)}\right) - \left(\varrho^2 + \frac{2 M_s^2}{H_{ss}^{\left(2\right)}}\right) \left(\frac{h_s}{M_s} - 2 h_{ss}^{\left(2\right)}\right)}
\end{align}

In the spin-1 sectors no mixing occurs between the neutral mesons, and the quadratic Lagrangians are simply written, in the isospin limit, as
\begin{align}
\mathcal{L}_V^{\left(mass\right)} & = \frac{1}{2} \rho^{0\mu} \rho^0_{\mu} 3 \varrho^2 H_{uu}^{\left(1\right)} + \frac{1}{2} \omega^{\mu} \omega_{\mu} 3 \varrho^2 H_{uu}^{\left(1\right)} + \frac{1}{2} \varphi^{\mu} \varphi_{\mu} 3 \varrho^2 H_{ss}^{\left(1\right)} + \rho^{+\mu}\rho^{-}_{\mu} \frac{3}{2} \varrho^2 H_{11}^{\left(1\right)} \nonumber \\
+ & \left(K^{*+\mu}K^{*-}_{\mu} + K^{0\mu}\bar{K}^{0}_{\mu}\right) \frac{3}{2} \left[ \varrho^2 H_{44}^{\left(1\right)} + \left(M_{u} - M_s\right)^2 \right],
\end{align}
and
\begin{align}
\mathcal{L}_A^{\left(mass\right)} & = \frac{1}{2} a_1^{0\mu} a^0_{1\mu} \left[ 3 \varrho^2 H_{uu}^{\left(2\right)} + 6 M_{u}^2 \right] + \frac{1}{2} f_1^{\mu} f_{1\mu} \left[ 3 \varrho^2 H_{uu}^{\left(2\right)} + 6 M_{u}^2 \right] \nonumber \\
& + \frac{1}{2} f_1^{'\mu} f'_{1\mu} \left[ 3 \varrho^2 {H_{ss}}^{\left(2\right)} + 6M_s^2 \right] + a_1^{+\mu}a^{-}_{1\mu} \left[ \frac{3}{2} \varrho^2 H_{11}^{\left(2\right)} + 6M_{u}^2 \right] \nonumber \\
& + \left(K_1^{*+\mu}K^{*-}_{1\mu} + K_1^{0\mu}\bar{K}^{0}_{1\mu} \right) \frac{3}{2} \left[ \varrho^2 H_{44}^{\left(2\right)} + \left(M_{u} + M_s\right)^2 \right],
\end{align}
where we have identified $\omega_{ns} \equiv \omega$, $\omega_s \equiv \varphi$, $f_{1ns} \equiv f_1$ and $f_{1s} \equiv f'_1$.

\begin{acknowledgments}
Work supported in part by Funda\c{c}\~{a}o para a Ci\^{e}ncia e a Tecnologia (FCT), PhD grant SFRH/BD/110315/2015, and Centro de F\'{i}sica da Universidade de Coimbra (CFisUC).
\end{acknowledgments}

\bibliography{article}

\end{document}